\documentclass[aps,%
twocolumn,%
floatfix,%
pra,%
amsfonts,%
groupedaddress,%
superscriptaddress]{revtex4-2}%
\usepackage[utf8]{inputenc}
\usepackage[T1]{fontenc}
\usepackage{mathrsfs}
\usepackage{mathtools}
\usepackage{amsmath}
\usepackage{amsfonts}
\usepackage{amssymb}
\usepackage[titletoc,title]{appendix}
\usepackage{algorithm,algorithmic}

\renewenvironment{algorithm}[1][t]{%
  \begin{figure}[#1]%
  \captionsetup{type=algorithm,name=Algorithm,labelformat=boldalgorithm}%
  \centering%
  \begingroup%
  \small%
  \setlength{\abovecaptionskip}{1pt}%
  \setlength{\belowcaptionskip}{1pt}%
  \let\algorithmoldcaption\caption%
  \renewcommand{\caption}[1]{%
    \algorithmoldcaption{##1}%
    \vspace{2pt}\hrule height 0.45pt\vspace{4pt}%
  }%
  \begin{minipage}{0.98\columnwidth}%
  \hrule height 0.65pt\vspace{2pt}%
}{%
  \vspace{2pt}\hrule height 0.65pt%
  \end{minipage}%
  \endgroup%
  \end{figure}%
}
\usepackage{bbm}
\usepackage{bm}
\usepackage{amsthm}
\usepackage{makecell}
\usepackage{array}
\usepackage[dvipsnames,table,xcdraw]{xcolor}
\definecolor{beamer@blendedblue}{rgb}{0.2,0.2,0.7}
\usepackage[osf,sc]{mathpazo}
\usepackage{caption}
\DeclareCaptionLabelFormat{boldalgorithm}{\textbf{Algorithm} #2}
\usepackage{subfig}
\usepackage{graphicx}
\usepackage{overpic}
\usepackage{booktabs}
\usepackage{multirow}
\usepackage[skins,breakable,most]{tcolorbox}
\usepackage{pgfplots}
\pgfplotsset{compat=1.18}
\usepgfplotslibrary{fillbetween}
\usetikzlibrary{arrows.meta,calc,positioning}
\newcolumntype{C}[1]{>{\centering\arraybackslash}p{#1}}
\usepackage[colorlinks=true,%
linkcolor=beamer@blendedblue,%
bookmarks=true,%
breaklinks=true,%
filecolor=beamer@blendedblue,%
anchorcolor=yellow,%
citecolor=beamer@blendedblue,%
urlcolor=beamer@blendedblue,%
pdfauthor={Kun Wang},%
pdfsubject={Adaptive Stabilizer State Fidelity Certification},%
CJKbookmarks=true]{hyperref}
\usepackage{float}
\usepackage{pifont}

\newtheorem{theorem}{Theorem}
\newtheorem{proposition}[theorem]{Proposition}

\newtheorem{corollary}[theorem]{Corollary}

\mathchardef\ordinarycolon\mathcode`\:
\mathcode`\:=\string"8000
\def\vcentcolon{\mathrel{\mathop\ordinarycolon}}
\begingroup \catcode`\:=\active
  \lowercase{\endgroup
  \let :\vcentcolon
  }

\tcolorboxenvironment{proof}{%
  enhanced,
  boxrule=0.5pt,%
  breakable,
  sharp corners
}

\DeclareFontFamily{U}{mathx}{\hyphenchar\font45}
\DeclareFontShape{U}{mathx}{m}{n}{<-> mathx10}{}
\DeclareSymbolFont{mathx}{U}{mathx}{m}{n}
\DeclareMathAccent{\widebar}{0}{mathx}{"73}
\DeclareMathOperator*{\argmax}{arg\,max}
\DeclareMathOperator*{\argmin}{arg\,min}

\newcommand{\wt}[1]{\widetilde{#1}}
\newcommand{\wh}[1]{\widehat{#1}}

\newcommand{\ket}[1]{\vert{#1}\rangle}

\newcommand{\ketbra}[1]{\vert{#1}\rangle\!\langle{#1}\vert}

\DeclareMathOperator{\tr}{Tr}
\newcommand{\1}{\mathbbm{1}}

\DeclareMathOperator{\rank}{rank}

\newcommand{\ba}{\bm{a}}
\newcommand{\bb}{\bm{b}}
\newcommand{\be}{\bm{e}}
\newcommand{\bs}{\bm{s}}
\newcommand{\bu}{\bm{u}}
\newcommand{\bv}{\bm{v}}

\makeatletter
\newsavebox{\@brx}
\newcommand{\llangle}[1][]{\savebox{\@brx}{\(\m@th{#1\langle}\)}%
  \mathopen{\copy\@brx\kern-0.5\wd\@brx\usebox{\@brx}}}
\newcommand{\rrangle}[1][]{\savebox{\@brx}{\(\m@th{#1\rangle}\)}%
  \mathclose{\copy\@brx\kern-0.5\wd\@brx\usebox{\@brx}}}
\makeatother

\newcommand*{\cF}{\mathcal{F}}

\newcommand*{\cH}{\mathcal{H}}

\newcommand*{\cP}{\mathcal{P}}

\newcommand{\bE}{\mathbb{E}}

\newcommand{\bP}{\mathbb{P}}
\newcommand{\bF}{\mathbb{F}}

\definecolor{wildstrawberry}{rgb}{1.0, 0.26, 0.64}
\definecolor{googleblue}{HTML}{4285F4}
\definecolor{googlered}{HTML}{DB4437}
\definecolor{googleyellow}{HTML}{F4B400}
\definecolor{googlegreen}{HTML}{0F9D58}

\newcommand{\Ftwo}{\bF_2}
\newcommand{\GL}{\mathrm{GL}}

\begin{document}

\newcommand{\thetitle}{{Adaptive Stabilizer State Fidelity Certification}}
\title{\thetitle}

\author{Kun Wang}
\email{nju.wangkun@gmail.com}
\affiliation{College of Computer Science and Technology, 
National University of Defense Technology, 
Changsha 410073, China}%

\date{\today}

\begin{abstract}
Certifying the fidelity of a prepared state to a target stabilizer state is a fundamental task 
in quantum information processing. 
Ref.~[\href{https://link.aps.org/doi/10.1103/PhysRevA.99.042337}{Phys. Rev. A \textbf{99}, 042337 (2019)}]
gave the optimal worst-case lower bound from one fixed stabilizer generator gauge,
but gauge dependence can leave a large fidelity ambiguity. We develop an adaptive
extension that reports the full certified fidelity interval. First, for a single gauge, we derive
the complementary optimal worst-case upper bound. 
We then formulate gauge selection as an adaptive design problem in which each round solves exact
endpoint linear programs and chooses a new gauge by a witness elimination policy. We prove
monotonic tightening, exact recovery once all nontrivial stabilizers are covered, and the worst-case
necessity of full coverage. Finally, we identify structured syndrome distributions for which
adaptivity beats this exponential benchmark, and we numerically confirm faster concentration.
\end{abstract}

\maketitle

\section{Introduction}
Stabilizer states are pure quantum states specified as the common $+1$ eigenspace of an
abelian Pauli group generated by independent commuting Pauli operators~\cite{Gottesman1997}.
This compact Pauli description makes them central to quantum information processing, with
quantum error correction (QEC)~\cite{Terhal2015} and measurement-based quantum computing
(MBQC)~\cite{RaussendorfBriegel2001} as two prominent examples. 
In QEC, stabilizer codes protect logical information through Pauli checks, 
while in MBQC, cluster states are stabilizer states that provide the entangled resource. 
Stabilizer state resources have been demonstrated on leading quantum
platforms and are advancing rapidly. Particularly, 
photonic platforms have demonstrated four-qubit cluster states~\cite{Walther2005};
superconducting processors have realized surface code and repetitive error detection
experiments~\cite{Barends2014,Kelly2015}, multipartite and cluster state
entanglement~\cite{Cao2023,Jiang2026}, topologically ordered stabilizer states~\cite{Satzinger2021},
and surface code logical qubits~\cite{GoogleQuantumAI2023}; trapped-ion platforms have
demonstrated fault tolerant control of logical qubits~\cite{Egan2021}; 
and neutral-atom arrays have enabled reconfigurable logical processors 
based on stabilizer codes~\cite{Bluvstein2024,Bluvstein2025}.

When such states are used as quantum resources, experimental progress must be accompanied by a
quantitative check of preparation quality: 
\emph{how close is the prepared state to the target stabilizer state?} 
This is the task of stabilizer fidelity estimation and certification.
Quantum state tomography can answer this question after reconstructing the state, 
but its measurement and classical postprocessing costs grow rapidly with system size, 
even though compressed sensing and tensor network variants reduce the cost under
additional structure~\cite{James2001,Altepeter2005,Gross2010,Cramer2010}. Stabilizer state
verification gives powerful hypothesis testing guarantees using Pauli or local
measurements~\cite{HayashiMorimae2015,DangniamHanZhu2020,LiHanZhu2020,ChenXieXuWang2025,ZhengYuZhangXuWang2026},
but it is formulated as accepting or rejecting a preparation under a prescribed test
strategy rather than as computing the tight fidelity interval implied by a fixed data set. 
Direct fidelity estimation is sample efficient for
stabilizer targets~\cite{FlammiaLiu2011,daSilvaLandonCardinalPoulin2011}, but it produces an
estimator rather than a worst-case fidelity certificate from the observed stabilizer expectations.
The closest precursor to this work is the fidelity certificate of Kalev, Kyrillidis, and
Linke~\cite{Kalev2019}, hereafter the KKL certificate, 
which measures one chosen stabilizer generator and returns the optimal worst-case lower fidelity 
consistent with those data. This certificate is experimentally friendly and adversarially meaningful,
but it provides only the lower endpoint and may leave large fidelity ambiguity
when different generator gauges reveal different stabilizer constraints.

\textbf{Contributions.}
We address these drawbacks by making the certificate \emph{adaptive} and \emph{interval-valued}.
First, we derive the complementary optimal worst-case upper bound, 
so that one chosen generator gauge yields a full certified fidelity interval.
Second, we develop an adaptive certified interval framework that treats the generator gauge as a design variable and uses a witness-elimination principle to select informative new gauges. 
Geometrically, the method progressively cuts down the feasible syndrome polytope; 
analytically and numerically, it tightens the certified interval and concentrates 
faster than uniform random gauge selection.

\textbf{Organization.} Sec.~\ref{sec:gauge-formulation} reviews stabilizer gauges and the KKL
certification task. Sec.~\ref{sec:drawbacks-KKL-certificate} explains two limitations of the fixed
gauge certificate. Sec.~\ref{sec:certified-upper-bound} gives the companion certified upper
endpoint from one chosen gauge. Sec.~\ref{sec:adaptive-framework} introduces the adaptive algorithm, and
Sec.~\ref{sec:performance-guarantees} proves its main guarantees. Sec.~\ref{sec:numerical-simulations}
presents the numerical simulations. The final sections compare with related methods and discuss
extensions.

\textbf{Notation and conventions.}
All binary linear algebra is over the finite field $\Ftwo$. 
For a positive integer $n$, write $[n]:=\{1,\cdots,n\}$. 
Let $\boldsymbol{1}:=(1,\cdots,1)\in\Ftwo^n$. 
Vectors $\bu,\bs\in\Ftwo^n$ are paired by the mod-$2$ inner product
\begin{align}
\bu\cdot \bs:=\sum_{j=1}^n u_js_j\pmod 2.
\end{align}
For a list of binary vectors, $\rank(\ba_1,\cdots,\ba_m)$ 
means the dimension of their span over $\Ftwo$.
For a finite set $\Omega$, $\Delta(\Omega)$ denotes the probability simplex on $\Omega$, and $\delta_x$ denotes the point mass at $x\in\Omega$. For $p\in\Delta(\Ftwo^n)$, we use the Walsh transform convention
\begin{align}
\wh p(\bu):=\sum_{\bs\in\Ftwo^n}(-1)^{\bu\cdot \bs}p(\bs),\qquad \bu\in\Ftwo^n.
\end{align}
The group of invertible $n\times n$ binary matrices is denoted by $\GL(n,2)$. When $A\in\GL(n,2)$, 
its columns are written as $\ba_1,\cdots,\ba_n$; labels are viewed as column vectors when multiplied by matrices. 
Let $\cP_n$ denote the $n$-qubit Pauli group.
We write $\mathbbm{1}[\cdot]$ for an indicator, and $\bP$ and $\bE$ for probability and expectation.

\section{Gauge freedom in stabilizer fidelity certification}
\label{sec:gauge-formulation}

This section reviews the stabilizer states, syndrome, gauge, and KKL certification task.

\subsection{Stabilizer states and generator gauge freedom}
\label{sec:stabilizer-state-formalism}

Let $\ket{\psi}$ be a stabilizer state on $n$ qubits with a fixed reference stabilizer group
\begin{align}\label{eq:stabilizer-generators}
S=\langle g_1,\cdots,g_n\rangle\subseteq \cP_n,
\end{align}
where $g_1,\cdots,g_n$ are independent commuting stabilizer generators.
For each $\bu=(u_1,\cdots,u_n)\in \Ftwo^n$, write
\begin{align}
 g(\bu):=\prod_{j=1}^n g_j^{u_j}\in S.
\end{align}
Thus each vector in $\Ftwo^n$ labels a unique stabilizer element of $\ket{\psi}$.
We call this fixed map $\bu\mapsto g(\bu)$, determined by the original
generators $g_1,\cdots,g_n$, the \emph{reference labeling}; $\bu$ is the
reference label of the stabilizer element $g(\bu)$.
The standard syndrome projectors are
\begin{align}\label{eq:syndrome-projectors}
\forall \bs\in\Ftwo^n,\quad
\Pi_{\bs}=\frac{1}{2^n}\prod_{j=1}^n\bigl(\1+(-1)^{s_j}g_j\bigr). 
\end{align}
Because the generators are independent and commuting, the operators
$\{\Pi_{\bs}\}_{\bs\in\Ftwo^n}$ are mutually orthogonal rank-one projectors resolving the identity. 
They project onto the simultaneous eigenspaces of the
chosen generators, and $g_j\Pi_{\bs}=\Pi_{\bs}g_j=(-1)^{s_j}\Pi_{\bs}$.
Consequently,
\begin{align}
g(\bu)=\sum_{\bs\in\Ftwo^n}(-1)^{\bu\cdot \bs}\Pi_{\bs}.
\end{align}

For a state $\rho$, define the \emph{syndrome distribution}:
\begin{align}
 p_\rho(\bs):=\tr\left[\rho\Pi_{\bs}\right],\qquad \bs\in\Ftwo^n.
\end{align}
The stabilizer fidelity of $\rho$ with respect to the stabilizer state $\ket{\psi}$ is simply the probability of the trivial syndrome $\bs = \boldsymbol{0}$,
\begin{align}
 F(\rho,\psi) := \tr\left[\rho\ketbra{\psi}\right] = \tr\left[\rho\Pi_{\boldsymbol{0}}\right] = p_\rho(\boldsymbol{0}).
\end{align}
On the other hand, taking the expectation of a stabilizer element $g(\bu)$ in $\rho$ gives, for every $\bu\in\Ftwo^n$,
\begin{align}
\mu_\rho(\bu):=\tr\left[\rho g(\bu)\right]=\sum_{\bs\in\Ftwo^n}(-1)^{\bu\cdot \bs}p_\rho(\bs)=\wh p_\rho(\bu).
\end{align}
Thus each stabilizer expectation of $\rho$ is exactly one Walsh character of the syndrome distribution. 


Let $A\in\GL(n,2)$ with columns $\ba_1,\cdots,\ba_n$. Since these columns form a basis of $\Ftwo^n$,
$A$ induces a stabilizer generator basis with respect to $\{g_1,\cdots,g_n\}$ via
\begin{align}\label{eq:associated-generator-basis}
 h_i=g(\ba_i),\qquad i=1,\cdots,n.
\end{align}
The elements $\{h_i\}_i$ are independent and generate the same stabilizer group $S$. A label $\bv\in\Ftwo^n$ in the induced basis $H(A)=(h_1,\cdots,h_n)$ corresponds to the stabilizer
\begin{align}
h(\bv) :=
\prod_{i=1}^n h_i^{v_i}
=g\!\left(\sum_{i=1}^n v_i\ba_i\right),
\end{align}
so the induced basis label $\bv$ represents the same stabilizer element as the reference label vector
$\bu=\sum_i v_i\ba_i=A\bv$.
Thus $A$ records the reference labels of the induced generators column by column:
its $i$th column $\ba_i$ is the reference label of $h_i$, and a general
induced basis label $\bv$ is converted to the reference label of the same
stabilizer element by multiplication by $A$.

Throughout this work, we call the choice of an invertible matrix $A\in\GL(n,2)$, or equivalently the induced generator
basis $H(A)$ together with its coordinate map back to the reference labels, a \emph{gauge}. The
\emph{gauge freedom} is the freedom to change this choice: replacing $A$ changes the binary
coordinates used to name the generators and their products, while leaving the stabilizer group $S$
and all physical observables unchanged.

\subsection{Stabilizer fidelity certification}
\label{sec:stabilizer-fidelity-certification}

We now recall the KKL certification setting~\cite{Kalev2019} in the gauge notation: one fixed
gauge determines the accessible generator expectation data, and the certificate is the worst-case
fidelity compatible with those data.

\textbf{Data access model.} A data acquisition round first chooses a gauge
$A\in\GL(n,2)$, equivalently the induced generator basis $H(A)$. The accessible experimental
summary consists of the expectation values of the $n$ generators in that chosen gauge,
\begin{align}
\mu_\rho(\ba_1),\cdots,\mu_\rho(\ba_n).
\end{align}
We idealize these expectation values as exact. This model does not assume access to retained
individual shot outcomes, nor does it assume that expectation values of products of the chosen
generators can be inferred as additional constraints.
This is the KKL data interface~\cite{Kalev2019} expressed in the present notation based on gauge
covariance, and it leads to the following certification task for a fixed gauge.

\textbf{KKL certification task.}
For a chosen gauge $A$ with columns $\ba_1,\cdots,\ba_n$, the KKL certification task~\cite{Kalev2019}
asks for the \emph{smallest} target fidelity compatible with the returned generator expectations.
Within the syndrome distribution language, this is the optimization problem
\begin{subequations}\label{eq:kkl-polytope-form}
\begin{align}
F_{\min}(A;\rho) := \min_{p\in\Delta(\Ftwo^n)}&\; p(\boldsymbol{0})\\
\text{s.t.}&\; \wh p(\ba_i)=\mu_\rho(\ba_i)\quad \forall i\in[n].
\end{align}
\end{subequations}
Because both the constraints and the objective depend on the unknown state $\rho$ through its syndrome
distribution, this distributional optimization is equivalent to minimizing over all quantum states
that could have produced the same expectation data from one gauge. 
Ref.~\cite{Kalev2019} solves this problem in
closed form, giving the optimal lower fidelity certificate from one gauge
\begin{align}
\label{eq:kkl-bound}
F_{\min}(A;\rho)=
\max\left\{0,\,
1-\frac12\sum_{i=1}^n\bigl(1-\mu_\rho(\ba_i)\bigr)
\right\}.
\end{align}
We call this quantity the KKL certificate. 
It is optimal for the data produced by the \emph{chosen} gauge $A$.

The importance of KKL certification is twofold. 
First, it uses only one fixed gauge. 
The experiment needs to estimate the expectations of the chosen generators, rather
than the ability to measure the full stabilizer group or adaptively choose many observables, 
which makes the data access model simple and experimentally friendly.
Second, it addresses a critical \emph{adversarial certification} scenario: 
among all quantum states consistent with the measured generator expectations, it reports
the smallest possible stabilizer fidelity, and therefore gives a rigorous worst-case lower fidelity
certificate without assuming a specific noise model. 

\section{Drawbacks of the KKL certificate}
\label{sec:drawbacks-KKL-certificate}

This section explains why the fixed-gauge KKL certificate is not the end of the story. First, it
reports the optimal lower endpoint from one chosen gauge, but the same generator data also define a
certified upper endpoint. This motivates the companion upper certificate in
Sec.~\ref{sec:certified-upper-bound}. Second, the lower certificate can depend sharply on the chosen
gauge, and additional gauges can remove fidelity ambiguity without measuring the full non-identity
stabilizer group.

\subsection{Gauge dependence and residual ambiguity}

From Eq.~\eqref{eq:kkl-bound}, the KKL certificate can depend strongly on the chosen gauge:
two gauges for the same stabilizer
group can yield different lower certificates for the same quantum state $\rho$. Sec.~\ref{sec:illustrative-examples}
gives explicit examples. Prop.~\ref{prop:singlebasis-insufficient} isolates the underlying information theoretic
reason. For any fixed gauge, the $n$ generator expectations need not uniquely determine $p(\boldsymbol{0})$.
\begin{proposition}
\label{prop:singlebasis-insufficient}
For every $A\in\GL(n,2)$ with columns $\ba_1,\cdots,\ba_n$ and every $n\ge 2$, there exist two syndrome distributions $p,q\in\Delta(\Ftwo^n)$ such that
\begin{align}
\wh p(\ba_i)=\wh q(\ba_i)\qquad \forall i\in[n],
\end{align}
but $p(\boldsymbol{0})\neq q(\boldsymbol{0})$.
\end{proposition}

The proof in Appx.~\ref{app:singlebasis-insufficient} constructs such $p$ and $q$ explicitly.
For any such pair, the two diagonal states
\begin{align}
\rho_p:=\sum_{\bs}p(\bs)\Pi_{\bs},
\qquad
\rho_q:=\sum_{\bs}q(\bs)\Pi_{\bs}
\end{align}
produce identical generator expectations in the chosen gauge $A$, but have different target fidelities. 
Thus data from a single gauge are not sufficient for fidelity determination in general.
We emphasize that Prop.~\ref{prop:singlebasis-insufficient} does \emph{not} contradict~\cite{Kalev2019}: 
the KKL certificate remains optimal for the data from one chosen gauge. 
Rather, Prop.~\ref{prop:singlebasis-insufficient} says that the data from a single gauge may be
insufficient to identify the fidelity. Additional stabilizer expectations are needed to remove this
residual ambiguity.

This motivates replacing fixed-gauge certification by \emph{adaptive} gauge choice: each new gauge adds
further Walsh character constraints,
shrinking the feasible set and tightening the certified fidelity interval.

\subsection{Illustrative examples}
\label{sec:illustrative-examples}

The following three qubit example gives compact motivation for adaptive certification by gauge
choice. Take the target state $\ket{\psi}=\ket{000}$ with reference generators $Z_1,Z_2,Z_3$, and
consider the diagonal state
\begin{align}
\rho_{\mathrm{ex}}
=\frac14\bigl(&\ketbra{000}+\ketbra{100}\nonumber\\
&+\ketbra{010}+\ketbra{001}\bigr),
\end{align}
whose true fidelity is $F(\rho_{\mathrm{ex}},\ket{\psi})=1/4$.

\paragraph*{Gauge choice matters.}
For the canonical gauge $A_{\mathrm{can}}$, whose induced generators are
$(Z_1,Z_2,Z_3)$, the measured expectations are $(1/2,1/2,1/2)$, and
Eq.~\eqref{eq:kkl-bound} is already exact, giving $L_{\mathrm{KKL}}=1/4$. For the equally valid
parity gauge $A_{\mathrm{par}}$, whose induced generators are
$(Z_1Z_2,\ Z_1Z_3,\ Z_1Z_2Z_3)$, the measured expectations are instead $(0,0,-1/2)$, and the same
KKL formula gives the trivial lower certificate $L_{\mathrm{KKL}}=0$. Thus two
gauges for the same stabilizer group and the same state can produce radically different KKL
certificates. The same one gauge data also carry upper endpoint information: the canonical gauge
gives the upper certificate $3/4$, while the parity gauge gives the upper certificate $1/4$. Thus
the upper endpoint can be informative even when the lower KKL certificate is loose.

\paragraph*{Additional gauges help.}
The parity gauge also shows why an additional gauge can be genuinely useful. If only
$A_{\mathrm{par}}$ is used, the data are compatible with a zero fidelity explanation, so the lower
certificate from one gauge is vacuous. Adding the canonical gauge $A_{\mathrm{can}}$ supplies the
missing single qubit stabilizer expectations,
\begin{align}
\wh p(\boldsymbol{100})=\wh p(\boldsymbol{010})=\wh p(\boldsymbol{001})=1/2.
\end{align}
Together with the parity gauge constraints, these data force the syndrome distribution of
$\rho_{\mathrm{ex}}$, and hence force the fidelity to be exactly $1/4$. However, the two queried
gauges cover only the labels
$\boldsymbol{100},\boldsymbol{010},\boldsymbol{001},\boldsymbol{110},\boldsymbol{101},\boldsymbol{111}$;
the stabilizer $Z_2Z_3$, corresponding to $\bu=\boldsymbol{011}$, is never queried. Thus the
improvement comes from a genuinely informative additional gauge, not from measuring the whole
non-identity stabilizer group.

\section{Certified upper bound}
\label{sec:certified-upper-bound}

The KKL certificate gives the optimal certified lower bound on the stabilizer fidelity 
from the generator data of a chosen gauge. 
We show that the same data also determine an optimal certified upper bound,
so one chosen gauge already certifies a full fidelity interval.
All proofs in this section are collected in Appx.~\ref{app:certified-upper-bound}.

Replacing the minimization in Eq.~\eqref{eq:kkl-polytope-form} by a maximization gives the
certified upper bound on the stabilizer fidelity for the same one gauge feasible set:
\begin{subequations}\label{eq:kkl-upper-polytope-form}
\begin{align}
F_{\max}(A;\rho) := \max_{p\in\Delta(\Ftwo^n)}&\; p(\boldsymbol{0})\\
\text{s.t.}&\; \wh p(\ba_i)=\mu_\rho(\ba_i)\quad \forall i\in[n].
\end{align}
\end{subequations}
We show that this problem admits an analytic solution.

\begin{proposition}\label{prop:single-gauge-upper}
The optimal value of Eq.~\eqref{eq:kkl-upper-polytope-form} is
\begin{align}\label{eq:kkl-upper-bound}
F_{\max}(A;\rho) = \frac{1}{2} + \frac{1}{2}\min_{i\in[n]}\mu_\rho(\ba_i).
\end{align}
\end{proposition}

Prop.~\ref{prop:single-gauge-upper}
complements the KKL certificate in Eq.~\eqref{eq:kkl-bound}. Together, the generator data
from one chosen gauge certify the full fidelity interval
\begin{align*}
\left[
\max\left\{0,1-\sum_{i=1}^n\frac{1-\mu_\rho(\ba_i)}{2}\right\},
\frac{1}{2} + \frac{1}{2}\min_{i\in[n]}\mu_\rho(\ba_i)
\right].
\end{align*}
This interval corresponds to the case of $t=1$ in Fig.~\ref{fig:adaptive-tightening}.

\textbf{Finite estimates.}
Experimentally, the expectation values of stabilizers from a chosen gauge can only 
be approximated from measurement statistics and therefore have shot noise.
The upper fidelity certificate also admits a direct finite-shot version.
For the generator labelled by $\ba_i$, let $Y_{i,j}\in\{0,1\}$ record the outcome of the $j$th measurement, with $Y_{i,j}=1$ for
the $+1$ outcome and $Y_{i,j}=0$ for the $-1$ outcome. Then
\begin{align}
a_i &:= \mathbb{E}[Y_{i,j}] = \frac{1+\mu_\rho(\ba_i)}{2}, \\
\widehat a_i &:= \frac{1}{m_i}\sum_{j=1}^{m_i}Y_{i,j} = \frac{1+\widehat\mu_i}{2}.
\end{align}
Define the empirical upper endpoint
\begin{align}
\widehat U_A:=\min_{i\in[n]}\widehat a_i
=\min_{i\in[n]}\frac{1+\widehat\mu_i}{2}.
\end{align}
Prop.~\ref{prop:single-gauge-upper-finite} gives a confidence guarantee for this empirical upper endpoint.

\begin{proposition}
\label{prop:single-gauge-upper-finite}
Fix a chosen gauge $A$, an accuracy parameter $\varepsilon>0$, and a failure probability
$\delta\in(0,1)$. If every generator in the gauge is measured independently at least
\begin{align}
\label{eq:upper-shot-per-generator}
m_i\ge \left\lceil \frac{\log(2n/\delta)}{2\varepsilon^2}\right\rceil
\end{align}
times, then with probability at least $1-\delta$,
\begin{align}
\label{eq:upper-finite-confidence}
\left|\widehat U_A-F_{\max}(A;\rho)\right|\le \varepsilon .
\end{align}
\end{proposition}

Since the true fidelity is never larger than the exact maximum compatible with the one gauge data,
Prop.~\ref{prop:single-gauge-upper-finite} immediately gives a certified upper fidelity bound 
on the stabilizer fidelity from finite samples.

\begin{corollary}
\label{cor:single-gauge-upper-sample}
Under the sampling rule in Eq.~\eqref{eq:upper-shot-per-generator}, the true stabilizer fidelity
obeys
\begin{align}
\label{eq:finite-upper-certificate}
F(\rho,\psi)\le \widehat U_A+\varepsilon
\end{align}
with probability at least $1-\delta$. Thus a chosen gauge upper fidelity certificate with additive
statistical slack $\varepsilon$ uses
\begin{align}
\label{eq:upper-total-shots}
M_{\mathrm{u}} = \sum_{i=1}^n m_i = n\left\lceil \frac{\log(2n/\delta)}{2\varepsilon^2}\right\rceil
\end{align}
copies of the state.
\end{corollary}

The finite-shot behavior differs from the KKL lower certificate because the upper certificate is a
minimum of the estimated success probabilities. A uniform error $\varepsilon$ in all estimated
$a_i$ changes the upper endpoint by at most $\varepsilon$, whereas the KKL lower certificate contains
a sum over all $n$ generator errors. Hence the statistical overhead for the upper certificate is
linear in $n$ up to logarithmic factors. This completes the one gauge certified interval:
Ref.~\cite{Kalev2019} gives the lower fidelity certificate, while Prop.~\ref{prop:single-gauge-upper} and
Cor.~\ref{cor:single-gauge-upper-sample} give the companion upper fidelity certificate
and its finite-shot version.

\begin{table*}[t]
\centering
\setlength{\tabcolsep}{0pt} 
\setlength\heavyrulewidth{0.3ex}  
\renewcommand{\arraystretch}{1.5} 
\begin{tabular}{@{}C{0.10\textwidth}C{0.20\textwidth}C{0.20\textwidth}C{0.20\textwidth}C{0.20\textwidth}@{}}
\toprule
\multirow{2}{*}{$p_{\mathrm{conf}}$} & \multicolumn{2}{c}{\textbf{Trapped-ion}} 
& \multicolumn{2}{c}{\textbf{Superconducting}} \\
 & lower certificate & upper certificate & lower certificate & upper certificate \\
\midrule
$0.999$ & $0.905$ & $1.000$ & $0.575$ & $0.878$ \\
$0.990$ & $0.913$ & $1.000$ & $0.585$ & $0.872$ \\
$0.900$ & $0.923$ & $1.000$ & $0.596$ & $0.865$ \\
\bottomrule
\end{tabular}
\caption{\raggedright
\textbf{One gauge fidelity certificates for the GHZ experiments of Ref.~\cite{Kalev2019}.}
The lower certificates are reported in~\cite[Table I]{Kalev2019}. The upper certificates are
computed from the same experimental frequencies using Eq.~\eqref{eq:kkl-upper-bound} and
Cor.~\ref{cor:single-gauge-upper-sample}. For each row, the corresponding one-sided statement holds
with probability at least $p_{\mathrm{conf}}$.}
\label{tab:kkl-ghz-upper}
\end{table*}

\textbf{Example using the KKL GHZ experiments.}
Ref.~\cite{Kalev2019} applied the lower certificate to three qubit GHZ states prepared on a trapped-ion
device and on the IBM superconducting platform. The measured frequencies are reported in
Figs.~3 and 4 of that work. For the trapped-ion data, the chosen generator expectations are
approximately $\mu(Z_1Z_2)=0.958$, $\mu(Z_2Z_3)=0.981$, and $\mu(X_1X_2X_3)=0.969$,
which gives $\widehat U_A=0.979$. For the IBM data, using the gauge
$(Z_1Z_2,Z_1Z_3,X_1X_2X_3)$ gives $\mu(Z_1Z_2)=0.771$, $\mu(Z_1Z_3)=0.817$, and $\mu(X_1X_2X_3)=0.676$,
and hence $\widehat U_A=0.838$. Applying Cor.~\ref{cor:single-gauge-upper-sample} with the same
effective per generator sample counts used in Table I of Ref.~\cite{Kalev2019}, namely
$11000/3$ for the trapped-ion data and $8192/3$ for the IBM data, gives the upper bounds in
Tab.~\ref{tab:kkl-ghz-upper}; the arithmetic is detailed in
Appx.~\ref{app:ghz-upper-calculation}. The lower columns reproduce the KKL lower certificates. Each
lower entry and each upper entry is a one-sided statement at confidence level $p_{\mathrm{conf}}$.

Tab.~\ref{tab:kkl-ghz-upper} should be read as a pair of complementary one-sided
certificates. For the trapped-ion experiment, the lower certificate already exceeds $0.90$ at all
listed confidence levels, while the upper certificate is clipped by the physical bound $1$; these
data certify a high fidelity preparation, but the chosen gauge and shot budget do not give a
nontrivial upper ceiling below $1$. For the IBM experiment, the same retained data certify a
moderate fidelity from below and also rule out fidelities above roughly $0.86$--$0.88$, depending on
the confidence level. If a simultaneous two-sided interval is desired, the two statements
can be combined by a union bound; for example, at $p_{\mathrm{conf}}=0.99$, the IBM data imply
$0.585\le F(\rho,\psi)\le 0.872$ with confidence at least $0.98$. Thus the upper certificate turns
the KKL analysis from a validation lower bound into a diagnostic interval: a wide interval indicates
that more shots or additional gauges are needed to localize the fidelity more sharply.

\section{Adaptive stabilizer fidelity certification}
\label{sec:adaptive-framework}
Sec.~\ref{sec:drawbacks-KKL-certificate} showed that a fixed gauge can leave substantial fidelity
ambiguity, while Sec.~\ref{sec:certified-upper-bound} turned one gauge data into a certified
interval with a KKL certified lower bound and a complementary certified upper bound. 
The next question is how to reduce this interval when more generator data can be acquired.

This section develops an adaptive answer. After each queried gauge, we keep all syndrome
distributions compatible with the accumulated Walsh constraints, report the certified fidelity
interval over this feasible set, and choose the next gauge to shrink the current ambiguity as much as possible. 

\subsection{Feasible polytope}\label{sec:feasible-polytope}

After $t$ rounds, suppose gauges $A_1,\cdots,A_t$ have been queried, where the columns of $A_r$ are $\ba_{r,1},\cdots,\ba_{r,n}$. In the ideal data access model, querying the gauge $A_r$ reveals the 
exact expectations $\mu_\rho(\ba_{r,1}),\cdots,\mu_\rho(\ba_{r,n})$. Let
\begin{align}
Q_t:=\{\ba_{r,i}: r\le t,\ i\in[n]\}
\end{align}
be the set of stabilizer labels queried up to round $t$. 
The exact \emph{feasible polytope} determined by the first $t$ rounds is
\begin{align}
\label{eq:feasible-polytope}
\cF_t:=\{p\in \Delta(\Ftwo^n):
\wh p(\bu)=\mu_\rho(\bu)\ \forall \bu\in Q_t\}.
\end{align}
Correspondingly, the certified \emph{fidelity interval} is given by
\begin{align}
\label{eq:certified-interval}
L_t:=\min_{p\in\cF_t} p(\boldsymbol{0}),\qquad U_t:=\max_{p\in\cF_t} p(\boldsymbol{0}),
\end{align}
where $L_t$ represents the lower fidelity certificate and $U_t$ represents the upper fidelity certificate
given by the $t$-th round. Define the \emph{interval width} $W_t=U_t-L_t$.
By construction, the true distribution belongs to $\cF_t$, so
\begin{align}
L_t\le F(\rho,\psi)\le U_t.
\end{align}
When $t=1$, this construction recovers the one gauge fidelity interval obtained in
Sec.~\ref{sec:certified-upper-bound}.

\subsection{Adaptive choice of the next gauge}

\begin{figure*}[t]
\centering
\includegraphics[width=0.92\textwidth]{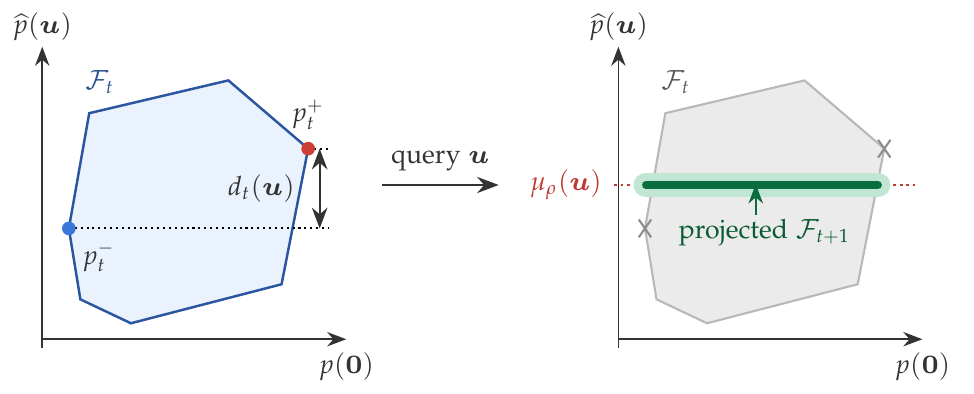}
\caption{\raggedright
\textbf{Geometric illustration of the witness elimination principle.} 
Left: the current feasible polytope $\cF_t$ projected onto the coordinates $p(\boldsymbol{0})$ and $\wh p(\bu)$; 
the endpoint witnesses attain the current fidelity interval endpoints and have vertical separation $d_t(\bu)$. 
Right: after querying $\bu$, the value $\mu_\rho(\bu)$ imposes the hyperplane $\wh p(\bu)=\mu_\rho(\bu)$. The green line segment is the image of the updated polytope $\cF_{t+1}=\cF_t\cap\{p:\wh p(\bu)=\mu_\rho(\bu)\}$ in this two coordinate projection; the full updated feasible set remains a polytope in the ambient syndrome probability space.}
\label{fig:hyperplane-cut}
\end{figure*}

The core adaptive question is how to choose $A_{t+1}$ so that we can further shrink the interval width. Let
\begin{align}\label{eq:endpoint-witnesses}
 p_t^-\in\argmin_{p\in\cF_t}p(\boldsymbol{0}),
 \qquad
 p_t^+\in\argmax_{p\in\cF_t}p(\boldsymbol{0})
\end{align}
be the lower and upper endpoint witnesses for the certified interval. These witnesses are feasible syndrome distributions that attain the two optimal values defining $[L_t,U_t]$. 
For any candidate stabilizer $\bu\in\Ftwo^n\setminus\{\boldsymbol{0}\}$, define its \emph{disagreement score}
\begin{align}
\label{eq:disagreement-score}
d_t(\bu):=\bigl|\wh p_t^+(\bu)-\wh p_t^-(\bu)\bigr|.
\end{align}
To make the acquisition policy in Eq.~\eqref{eq:endpoint-witnesses} reproducible, 
we can fix once and for all an ordering
$\bs_1,\cdots,\bs_{2^n}$ of $\Ftwo^n$. If either endpoint LP has multiple optimal
solutions, choose the lexicographically smallest probability vector on the corresponding optimal face. 
Equivalently, one may solve a fixed sequence of secondary linear optimizations over the optimal face. This convention
determines $p_t^-$, $p_t^+$, and hence $d_t$, without changing the certified endpoints
$L_t$ and $U_t$.

\textbf{The witness elimination principle.}
The key design idea behind the disagreement score metric is the \emph{witness elimination principle}. 
The current interval width $W_t$ is certified by two feasible endpoint witnesses, $p_t^-$ and $p_t^+$, 
which agree with all data collected so far but assign different values to the target coordinate $p(\boldsymbol{0})$. 
Querying a new stabilizer label $\bu$ adds the linear constraint $\wh p(\bu)=\mu_\rho(\bu)$, 
i.e., it cuts the feasible polytope by a hyperplane. If $\wh p_t^-(\bu)$ and $\wh p_t^+(\bu)$ are far apart, 
then the two current endpoint witnesses make strongly different predictions for the outcome of this query. 
Once the true value $\mu_\rho(\bu)$ is revealed, at least one of these two particular witnesses is eliminated unless both predictions coincide. Thus $d_t(\bu)$ measures how directly the query tests the witnesses responsible for the present fidelity ambiguity. By contrast, if $d_t(\bu)=0$, then this query does not separate the two current endpoint witnesses and is not expected to reduce the interval in the direction of $p(\boldsymbol{0})$, although it may still remove other feasible points.
This principle is the design criterion behind the disagreement score and the subsequent gauge selection policy.
This geometry is summarized in Fig.~\ref{fig:hyperplane-cut}: in the two-dimensional projection onto the target coordinate $p(\boldsymbol{0})$ and the queried Walsh coordinate $\wh p(\bu)$, the before-query region $\cF_t$ and the after-query section $\cF_{t+1}$ can be seen separately, while $d_t(\bu)$ is the separation between the two endpoint witnesses in the queried direction. The updated feasible set remains a polytope in the ambient syndrome-probability space; it appears as a line segment in this projection only because the plotted coordinate $\wh p(\bu)$ has been fixed by the new query.

\textbf{Disagreement score of a gauge.}
A candidate gauge $A$ with columns $\ba_1,\cdots,\ba_n$ is then scored by
\begin{align}
\label{eq:basis-score}
\Phi_t(A):=\sum_{i=1}^n d_t(\ba_i).
\end{align}
This additive score is a tractable surrogate for the ideal but more expensive one step lookahead
objective that would recompute the fidelity interval for every possible next gauge and every
possible returned data vector. In the generator-only interface, one round returns the $n$ Walsh
coefficients associated with the columns of one gauge. Summing $d_t(\ba_i)$ therefore favors gauges
whose generators collectively test many high disagreement directions, while the independence
constraint prevents spending a round on a linearly dependent column set. The additivity also has an
algorithmic advantage: it converts gauge selection into the maximum weight problem over a vector
matroid below, so the next gauge can be found greedily rather than by enumerating $\GL(n,2)$. We use
$\Phi_t$ in this sense: not as a proof of optimal one step interval contraction, but as a
certificate driven acquisition policy designed to attack the current fidelity witnesses with a
computationally controlled gauge choice.

To avoid an entirely redundant round, the next gauge is chosen from gauges containing at least one 
previously unqueried stabilizer label. Specifically, for a gauge $A$ with column set
$C(A):=\{\ba_1,\cdots,\ba_n\}$, define
\begin{align}
\Gamma_t:=\{A\in\GL(n,2): C(A)\not\subseteq Q_t\}.
\end{align}
Thus every $A\in\Gamma_t$ contains at least one stabilizer label that has not
previously been queried. In a nonterminal round, the next gauge is chosen as
\begin{align}
\label{eq:next-basis}
A_{t+1}\in\argmax_{A\in\Gamma_t}\Phi_t(A).
\end{align}
We refer to the deterministic gauge selection rule in Eq.~\eqref{eq:next-basis} as the
\emph{witness elimination policy}.

\textbf{Converting to a maximum weight problem.}
Because the score in Eq.~\eqref{eq:basis-score} is additive over the columns of $A$, the column set
version of Eq.~\eqref{eq:next-basis} is a non-redundant maximum weight problem over a vector matroid
on $\Ftwo^n$. The ground set is $\Ftwo^n\setminus\{\boldsymbol{0}\}$, with weight $d_t(\bu)$ assigned to element
$\bu\in\Ftwo^n\setminus\{\boldsymbol{0}\}$. A feasible set is any linearly independent subset of
$\Ftwo^n\setminus\{\boldsymbol{0}\}$, and a vector matroid basis is any set
$B=\{\bb_1,\cdots,\bb_n\}\subset \Ftwo^n\setminus\{\boldsymbol{0}\}$ with $\rank(B)=n$. Thus the column set of the
next gauge is obtained by solving
\begin{align}
\label{eq:max-weight-vector-matroid}
B_{t+1}\in
\argmax_{\substack{B\subseteq \Ftwo^n\setminus\{\boldsymbol{0}\}\\
|B|=n,\ \rank(B)=n\\ B\not\subseteq Q_t}}
\sum_{\bu\in B}d_t(\bu).
\end{align}
Any ordering of the elements of $B_{t+1}$ gives the columns of a matrix $A_{t+1}\in\Gamma_t$, and
Eq.~\eqref{eq:max-weight-vector-matroid} is equivalent to Eq.~\eqref{eq:next-basis}.
By the standard greedy theorem for matroids, sorting elements by nonincreasing weight and adding an
element whenever it preserves independence returns a maximum weight basis for the underlying vector
matroid. Operationally, this greedy vector matroid scan computes the desired column set in every
nonterminal round. Indeed, if $\bu\in Q_t$, then every $p\in\cF_t$ has the same Walsh coefficient
$\wh p(\bu)=\mu_\rho(\bu)$, so $d_t(\bu)=0$. If $W_t>0$, the endpoint witnesses $p_t^-$ and $p_t^+$
must disagree on at least one unqueried nonzero Walsh coefficient; otherwise Walsh--Hadamard
inversion would force $p_t^-=p_t^+$ and hence $W_t=0$. Therefore some unqueried label has positive
weight, and any maximum weight vector matroid solution has positive total score and cannot be
contained in $Q_t$. One may thus sort all nonzero labels $\bu$ in nonincreasing order of
$d_t(\bu)$, scan this list, add $\bu$ exactly when it increases the rank of the selected set, and
stop once $n$ independent vectors have been selected. Zero weight ties should be broken in favor of
unqueried labels, which adds extra constraints at no loss in the acquisition score.

\subsection{The certification algorithm} 

Now we are ready to present the adaptive stabilizer fidelity certification algorithm.
The algorithm keeps the full set of syndrome distributions compatible with the stabilizer
expectations observed so far, and reports the smallest and largest possible values of $p(\boldsymbol{0})$ over
this feasible set. To choose the next gauge, it compares the lower and upper endpoint witnesses that
attain these two values. A stabilizer whose Walsh coefficient differs strongly between these
witnesses is expected to cut through the current ambiguity, so the next gauge is chosen to collect
$n$ independent high disagreement stabilizers. 
The formal procedure is described in Alg.~\ref{alg:adaptive}.

\begin{algorithm}[t]
\captionsetup{justification=raggedright,singlelinecheck=false}
\caption{Adaptive stabilizer fidelity certification}
\label{alg:adaptive}
\begin{tabular}{@{}l@{\hspace{0.35em}}>{\raggedright\arraybackslash}p{0.82\linewidth}@{}}
\textbf{Input:} & Target stabilizer state $\ket{\psi}$; unknown state $\rho$; threshold $\varepsilon\ge 0$.\\
\textbf{Output:} & A sequence of certified intervals $[L_t,U_t]$ for $F(\rho,\psi)$.
\end{tabular}
\vspace{-4pt}
\begin{algorithmic}[1]
\STATE Fix reference generators $g_1,\cdots,g_n$ for $\ket{\psi}$.
\STATE Choose an initial gauge $A_1\in\GL(n,2)$ and set $Q_0\gets\emptyset$.
\FOR{$t=1,2,\cdots$}
    \STATE Query gauge $A_t$ with columns $\ba_{t,1},\cdots,\ba_{t,n}$.
    \STATE Obtain $\mu_\rho(\ba_{t,i})$ for all $i\in[n]$.
    \STATE Set $C_t\gets\{\ba_{t,i}:i\in[n]\}$ and $Q_t\gets Q_{t-1}\cup C_t$.
    \STATE Compute the feasible polytope $\cF_t$ via Eq.~\eqref{eq:feasible-polytope}.
    \STATE Compute the two endpoint witnesses Eq.~\eqref{eq:endpoint-witnesses}.
    \STATE Set $L_t\gets p_t^-(\boldsymbol{0})$, $U_t\gets p_t^+(\boldsymbol{0})$, and $W_t\gets U_t-L_t$.
    \STATE Record the certified interval $[L_t,U_t]$.
    \IF{$W_t\leq \varepsilon$}
        \RETURN All recorded certified intervals.
    \ENDIF
    \STATE Compute $d_t(\bu)\gets|\wh p_t^+(\bu)-\wh p_t^-(\bu)|$ for every nonzero label $\bu$.
    \STATE Form $\Gamma_t=\{A\in\GL(n,2):C(A)\not\subseteq Q_t\}$.
    \STATE Choose $A_{t+1}\in\argmax_{A\in\Gamma_t}\Phi_t(A)$ via Eq.~\eqref{eq:next-basis}.
\ENDFOR
\end{algorithmic}
\end{algorithm}

\begin{figure*}[t]
\centering
\includegraphics[width=0.92\textwidth]{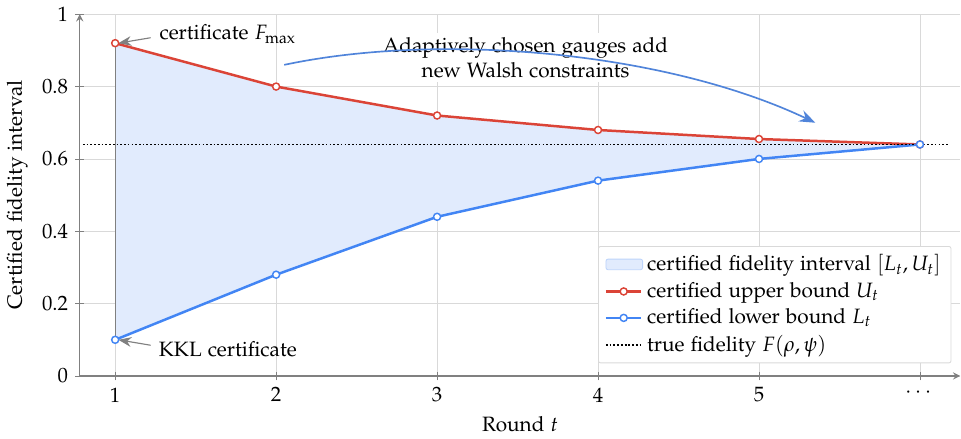}
\caption{\raggedright 
\textbf{Concentration of adaptive certified fidelity intervals.}
At the first round, the lower endpoint $L_1$ is exactly the KKL certificate in Eq.~\eqref{eq:kkl-bound}, 
while the upper endpoint $U_1$ is the companion upper certificate in Eq.~\eqref{eq:kkl-upper-bound}. 
In later rounds, Alg.~\ref{alg:adaptive} queries gauges whose labels are selected by the witness elimination
principle. The added Walsh constraints shrink the feasible polytope, so the lower endpoint $L_t$
can increase, the upper endpoint $U_t$ can decrease, and the certified interval $[L_t,U_t]$
concentrates around the true fidelity $F(\rho,\psi)$.}
\label{fig:adaptive-tightening}
\end{figure*}

The local geometry of each update is the cutting picture in Fig.~\ref{fig:hyperplane-cut}. A queried
label $\bu\notin Q_t$ imposes the new constraint $\wh p(\bu)=\mu_\rho(\bu)$, slicing the current
feasible polytope and testing the endpoint witnesses that certify the present ambiguity.
Fig.~\ref{fig:adaptive-tightening} shows the cumulative effect at the level of the reported
certificate. As adaptive rounds add new Walsh constraints, the feasible sets are nested, the lower
certified bound $L_t$ is nondecreasing, the upper certified bound $U_t$ is nonincreasing, and the
interval $[L_t,U_t]$ concentrates around the true fidelity.

\textbf{Algorithm variant.}
The witness elimination policy admits a finer single stabilizer version if the measurement
interface allows arbitrary stabilizer observables to be queried one at a time. That stronger access
variant, together with its monotonicity, completeness, and width rate guarantees, is given in
Appx.~\ref{app:fine-grained-certification-algorithm}. 
We keep Alg.~\ref{alg:adaptive} as the main protocol because it
matches the KKL generator-only interface and is the setting in which the $\GL(n,2)$ gauge freedom is
the central design resource. 

\section{Performance analysis}
\label{sec:performance-guarantees}
We now carry out performance analysis of Alg.~\ref{alg:adaptive}. 
All proofs in this section are deferred to Appx.~\ref{app:performance-guarantees}.

\subsection{Classical computational complexity}
\label{subsec:classical-complexity}
The exact implementation of Alg.~\ref{alg:adaptive} has three classical components in each
nonterminal round: 
1) solving the two endpoint linear programs, 
2) computing the disagreement spectrum, and
3) selecting the next gauge. 
We account for these components separately and then add their costs to obtain 
the total classical complexity.

After $t$ rounds, the number of queried nontrivial stabilizer labels satisfies
\begin{align}
q_t:=|Q_t|\le \min\{tn,2^n-1\}.
\end{align}
At round $t$, $\cF_t$ has one variable $p(\bs)$ for each syndrome
$\bs\in\Ftwo^n$, so it has $2^n$ variables and $O(2^n+q_t)$ linear constraints, including
nonnegativity, normalization, and the queried Walsh constraints.

\emph{1) Endpoint LP cost.}
The lower and upper endpoint problems are two linear programs over this syndrome vector.
A generic LP solver therefore has classical cost
\begin{align}
C_{t,\mathrm{LP}}
=\operatorname{poly}\!\left(2^n,q_t,\log(1/\tau_{\rm num})\right)
=2^{O(n)},
\end{align}
where $\tau_{\rm num}$ denotes the requested numerical accuracy. We do not assume a specialized
linear time LP algorithm; the polynomial degree can depend on the solver and numerical model.

\emph{2) Disagreement spectrum cost.}
After the endpoint witnesses $p_t^-$ and $p_t^+$ are obtained, the disagreement spectrum is the
Walsh transform of the length $2^n$ vector $p_t^+-p_t^-$. A fast Walsh--Hadamard transform gives
\begin{align}
C_{t,\mathrm{dis}}=O(n2^n).
\end{align}

\emph{3) Gauge selection cost.}
The next gauge is obtained by the greedy vector matroid algorithm in Eq.~\eqref{eq:max-weight-vector-matroid}.
It scans the $2^n-1$ nonzero stabilizer labels and performs rank updates over $\Ftwo$, giving
\begin{align}
C_{t,\mathrm{gauge}}=O\!\left(2^n\operatorname{poly}(n)\right),
\end{align}
where the precise polynomial factor depends on the implementation of the rank updates.

Combining the components gives the per round cost
\begin{align}
C_t =
\operatorname{poly}\!\left(2^n,q_t,\log(1/\tau_{\rm num})\right)
 + O\!\left(2^n\operatorname{poly}(n)\right).
\end{align}
For $T$ rounds,
\begin{align}
\label{eq:exact-total-complexity}
C_T = \sum_{t=1}^{T} C_t \sim O(2^n\operatorname{poly}(n)).
\end{align}
The classical memory required to store explicit syndrome probability vectors and their Walsh transforms is
$O(2^n)$. Thus the exact implementation is exponential in $n$. The memory and non-LP
postprocessing scale as $O(2^n\operatorname{poly}(n))$, while the endpoint LPs require
$\operatorname{poly}(2^n,q_t,\log(1/\tau_{\rm num}))$ time with generic solvers. The matroid
formulation is still important because it keeps gauge selection at the same
$O(2^n\operatorname{poly}(n))$ scale and avoids enumerating ordered gauges in $\GL(n,2)$.

\textbf{Endpoint optimization relaxation.}
The exponential dependence is the main computational limitation of the exact formulation. The
theorems and numerical simulations in this work use the exact optimization layer so that the
reported intervals are the tight certificates for the acquired data. For larger systems, the same
certification layer could be paired with a scalable relaxation of the endpoint optimization. To
remain certified, such a relaxation should optimize over an outer feasible set
$\widetilde{\cF}_t\supseteq\cF_t$, giving
\begin{align}
\widetilde L_t:=\min_{p\in\widetilde{\cF}_t}p(\boldsymbol{0})
\le F(\rho,\psi)\le
\max_{p\in\widetilde{\cF}_t}p(\boldsymbol{0})=:\widetilde U_t .
\end{align}
If the relaxed feasible sets are nested as new gauge constraints are added, the relaxed interval is
monotone by the same argument as Prop.~\ref{prop:monotonicity} below.

Possible scalable optimization layers include cutting plane or column generation methods with
certified separation~\cite{Kelley1960,BertsimasTsitsiklis1997,Desaulniers2005,LuebbeckeDesrosiers2005}. 
Other possibilities include sparse support or coarse grained syndrome relaxations with verified residual
mass bounds, tensor network surrogates with rigorous truncation control, and linear or semidefinite
moment relaxations that exploit low weight or geometrically local stabilizer structure. These
directions are not part of the main results here, and we do not numerically validate them in this
work. Their role is to indicate how the exact benchmark in Eq.~\eqref{eq:exact-total-complexity}
could be replaced by conservative optimization layers in future large-scale implementations.

\subsection{Monotonicity}
\label{subsec:monotonicity}
We first record the basic nesting property of the certified interval in the ideal data access
model. Querying a new gauge adds Walsh constraints without removing any old constraints, so
the feasible set can only shrink. 

\begin{proposition}
\label{prop:monotonicity}
The feasible sets are nested:
\begin{align}
\cF_{t+1}\subseteq \cF_t.
\end{align}
Consequently, $L_t$ is nondecreasing, $U_t$ is nonincreasing, and $W_t$ is nonincreasing in $t$.
\end{proposition}

\textbf{Full stabilizer coverage.}
Monotonicity explains why the certified interval can only tighten. The corresponding endpoint
statement below says that, once the queried gauges have covered all nontrivial stabilizer labels, the
feasible set has no remaining freedom and the interval collapses to the true fidelity. The missing
label $\bu=\boldsymbol{0}$ need not be queried, because $\wh p(\boldsymbol{0})=1$ for every
probability distribution.

\begin{proposition}
\label{prop:finite-completeness}
Suppose that for some round $t_\star$,
\begin{align}
Q_{t_\star}=\Ftwo^n\setminus\{\boldsymbol{0}\}.
\end{align}
Then, for every $t\ge t_\star$,
\begin{align}
\cF_t=\{p_\rho\},
\qquad
L_t=U_t=F(\rho,\psi).
\end{align}
Consequently, if the sequence of queried labels satisfies
\begin{align}
\bigcup_{t\ge 1}Q_t=\Ftwo^n\setminus\{\boldsymbol{0}\},
\end{align}
then $L_t$ and $U_t$ converge in finite time to the true fidelity.
\end{proposition}

Prop.~\ref{prop:monotonicity} and Prop.~\ref{prop:finite-completeness}
together give the formal justification for the qualitative behavior shown in Fig.~\ref{fig:adaptive-tightening}: 
the certified interval moves monotonically inward and becomes exact at full stabilizer coverage.

\textbf{Worst case limitation.}
The preceding proposition is a completeness statement: full stabilizer coverage is sufficient for
exactness. In the unrestricted state space, however, full coverage is also necessary in the
worst-case setting. 
Thus the interval in Fig.~\ref{fig:adaptive-tightening} is
guaranteed to move monotonically inward and to collapse after full coverage, but the collapse can
require exponentially many stabilizer labels for adversarial states.

\begin{proposition}
\label{prop:query-complexity-lower-bound}
Consider any adaptive generator-only protocol that always outputs the exact fidelity $F(\rho,\psi)$. 
Then, in the worst-case setting, the protocol must query all $2^n-1$ nontrivial stabilizers $g(\bu)$ with $\bu\in\Ftwo^n\setminus\{\boldsymbol{0}\}$. Consequently, exact fidelity determination 
for arbitrary states requires at least
\begin{align}
\left\lceil\frac{2^n-1}{n}\right\rceil
\end{align}
full gauge rounds in the worst-case setting.
\end{proposition}

\subsection{Random gauges and coverage}
\label{subsec:random-gauges-coverage}
Prop.~\ref{prop:query-complexity-lower-bound} rules out a universal fast certification guarantee for
arbitrary states, but it is still useful to quantify conservative rates before full coverage. One
might try to prove such a rate directly for the witness elimination policy in Alg.~\ref{alg:adaptive}.
However, its selection criterion is endpoint disagreement, not uniform coverage of all missing labels:
it can keep concentrating on a small set of high disagreement directions, while giving no 
\emph{state-independent} lower bound on the probability of querying an arbitrary missing label. 
Thus a universal coverage argument for the witness elimination policy would require assumptions on the
state or on how disagreement is distributed over the unqueried labels.

We therefore separate two kinds of guarantees. The following random gauges provide a 
\emph{conservative coverage baseline}: each missing label has a controlled probability of appearing, so the
remaining ambiguity can be bounded without any assumption on the state. The adaptive witness
elimination policy is analyzed separately through the endpoint-disagreement bounds, 
which show that it targets the current ambiguity but do not by themselves give a uniform coverage rate. 

\textbf{Randomized gauge policy.}
A \emph{uniform random gauge} means a matrix $A\in\GL(n,2)$ sampled uniformly from
the finite group $\GL(n,2)$. Equivalently, its columns are an ordered random basis of $\Ftwo^n$:
the first column is a uniformly random nonzero label, and each later column is sampled uniformly
outside the span of the previous columns. Since the algorithm records only the queried stabilizer
labels, the relevant object is the unordered column set of this random basis. By symmetry, for any
fixed nonzero label $\bu$,
\begin{align}
\label{eq:uniform-random-gauge-inclusion}
\bP\!\left(\bu \text{ is a column of } A\right)=\frac{n}{2^n-1}.
\end{align}
More generally, a \emph{randomized gauge policy} is any history-dependent procedure that samples
$A_{t+1}\in\GL(n,2)$ from a probability distribution determined by the current history $\cH_t$.
The \emph{uniform random gauge policy} samples a uniform random gauge at each round, independently
of the history.

The following Prop.~\ref{prop:coverage-rate} gives a \emph{state-independent}
coverage benchmark. It bounds the largest possible remaining interval using only the number of
nontrivial stabilizer labels not yet queried. It also gives an expected decay rate for $W_t$ under
any randomized gauge policy that gives every missing label a uniform chance of being included in
the next gauge.

\begin{proposition}
\label{prop:coverage-rate}
Let $m_t:=|(\Ftwo^n\setminus\{\boldsymbol{0}\})\setminus Q_t|$ be the number of nontrivial stabilizer labels not yet queried after round $t$. For every state $\rho$ and every sequence of queried gauges,
\begin{align}
\label{eq:coverage-width-bound}
W_t\le \min\left\{1,\frac{2m_t}{2^n}\right\}.
\end{align}
Moreover, suppose the gauge choices follow a randomized gauge policy, set $Q_0=\emptyset$, and assume that for some $\alpha\in[0,1]$, every unqueried label has conditional probability at least $\alpha$ of being included in the next queried gauge:
\begin{align}
\label{eq:coverage-probability}
\bP\!\left(\bu\in Q_{t+1}\mid \cH_t\right)\ge \alpha,
\qquad \bu\in (\Ftwo^n\setminus\{\boldsymbol{0}\})\setminus Q_t,
\end{align}
where $\cH_t$ is the history through round $t$. Then
\begin{align}
\label{eq:expected-width-rate}
\bE[W_t]\le \frac{2^n-1}{2^{n-1}}(1-\alpha)^t.
\end{align}
In particular, under the uniform random gauge policy, $\alpha=n/(2^n-1)$ and
\begin{align}
\label{eq:uniform-random-width-rate}
\bE[W_t]
\le \frac{2^n-1}{2^{n-1}}\left(1-\frac{n}{2^n-1}\right)^t
\le 2\exp\!\left(-\frac{nt}{2^n-1}\right).
\end{align}
\end{proposition}

Prop.~\ref{prop:coverage-rate} is deliberately conservative. It treats all unqueried labels as
potentially important and therefore gives a universal guarantee, but the uniform random rate has the
small exponent $n/(2^n-1)$. On the other hand, the deterministic witness elimination policy uses more information: it
looks at the two endpoint witnesses that currently certify the interval width and asks which
unqueried labels distinguish them best. The following quantities measure this remaining endpoint
disagreement:
\begin{align}
\begin{aligned}
D_t&:=\sum_{\bu\in(\Ftwo^n\setminus\{\boldsymbol{0}\})\setminus Q_t}d_t(\bu),\\
\Delta_t&:=\max_{\bu\in(\Ftwo^n\setminus\{\boldsymbol{0}\})\setminus Q_t}d_t(\bu).
\end{aligned}
\end{align}

\begin{proposition}
\label{prop:disagreement-mass}
For every nonterminal round of Alg.~\ref{alg:adaptive},
\begin{align}
\label{eq:disagreement-mass-bound}
W_t\le 2^{-n}D_t\le \frac{m_t}{2^n}\Delta_t .
\end{align}
Moreover, the witness elimination gauge chosen by Eq.~\eqref{eq:next-basis} satisfies
\begin{align}
\label{eq:gauge-score-lower-bound}
\Phi_t(A_{t+1})\ge \Delta_t\ge \frac{2^n}{m_t}W_t .
\end{align}
\end{proposition}

Prop.~\ref{prop:disagreement-mass} turns the LP endpoint gap into an actionable query signal. Because
the endpoint witnesses already agree on all queried Walsh coefficients, any remaining width must be
carried by unqueried labels; hence a nontrivial $W_t$ forces nontrivial endpoint disagreement among
those labels. The witness elimination policy then chooses a gauge whose score is at least the largest
such single-label disagreement, so the next gauge is guaranteed to target the labels currently
responsible for the certified gap. This is a deterministic targeting guarantee, not a uniform
coverage guarantee.

\textbf{Mixed randomized gauge policy.}
The deterministic targeting statement in Prop.~\ref{prop:disagreement-mass} 
is not the same as the coverage condition in Eq.~\eqref{eq:coverage-probability}. 
In particular, the witness elimination policy may repeatedly focus on the same
region of endpoint disagreement and need not give every still-unqueried label a uniform chance of
being tested. To obtain a state-independent rate while retaining the adaptive choice, 
we adopt a mixed randomized gauge policy: 
each nonterminal round queries one gauge, but the gauge is sampled from a mixture of 
the gauge selected by the witness elimination policy and a uniform random gauge.
Fix $\gamma\in[0,1]$. In each nonterminal round, use the witness elimination gauge with probability
$1-\gamma$ and sample a uniform random gauge in the sense of
Eq.~\eqref{eq:uniform-random-gauge-inclusion} with probability $\gamma$. Then every missing label
has conditional inclusion probability at least $\gamma n/(2^n-1)$, coming from the uniform-random
branch of this policy. The witness elimination branch may add further labels, and by monotonicity it
cannot increase the certified width. Therefore Prop.~\ref{prop:coverage-rate} applies with
$\alpha=\gamma n/(2^n-1)$. Starting from $Q_0=\emptyset$, we have
\begin{subequations}\label{eq:mixed-randomized-policy-rate}
\begin{align}
\bE[W_t]
&\le \frac{2^n-1}{2^{n-1}}\left(1-\frac{\gamma n}{2^n-1}\right)^t \\
&\le 2\exp\!\left(-\frac{\gamma nt}{2^n-1}\right).
\end{align}    
\end{subequations}

Taken together, Prop.~\ref{prop:coverage-rate}, Prop.~\ref{prop:disagreement-mass}, and the mixed
randomized gauge policy bound in Eq.~\eqref{eq:mixed-randomized-policy-rate} separate two kinds of statements. 
The uniform random gauge policy gives a universal but slow worst-case benchmark. 
The witness elimination gauge policy gives a deterministic certificate that the next
gauge chosen by the witness elimination policy targets the current endpoint ambiguity. 
The mixed randomized gauge policy combines both:
the uniform random branch supplies a universal convergence rate, while the witness elimination
branch obtains stronger contraction on structured instances.

\subsection{Affine-support syndrome states}\label{sec:affine-support-syndrome-states}

Props.~\ref{prop:finite-completeness} and~\ref{prop:query-complexity-lower-bound}
establish the full coverage benchmark: querying all $2^n-1$ nontrivial stabilizer labels is
sufficient for exact certification and is necessary in the worst-case setting over unrestricted states.
Here we explore structured states for which this exponential barrier can be avoided, so that the
adaptive certification protocol converges quickly. We identify structured syndrome distributions for
which the fidelity is determined by a much smaller set of Walsh characters. 

Let $\bs_0\in\Ftwo^n$ and $V\le \Ftwo^n$ be a linear subspace of dimension $r$. 
Define the probability distribution
\begin{align}
\forall\bs\in\Ftwo^n,\quad
 p_{\bs_0,V}(\bs):=2^{-r}\mathbbm{1}[\bs\in \bs_0+V].
\end{align}
A state $\rho$ is called an $r$-dimensional \emph{affine-support syndrome state} relative to
$\ket{\psi}$ if its syndrome distribution satisfies $p_\rho=p_{\bs_0,V}$ for some affine subspace
$\bs_0+V$. This name emphasizes the relevant state structure: the
syndrome support is an affine subspace, and the distribution is flat on that support.
This class of states is experimentally meaningful from a noisy preparation perspective:
an experimenter may aim to prepare $\ket{\psi}$, but the preparation can be followed by a classical
random Pauli error whose syndrome is sampled uniformly from the affine set $\bs_0+V$. 
The resulting state is then an affine-support syndrome state.

For this class, the useful object is not the full collection of Walsh characters but the annihilator
$V^\perp$. Indeed,
\begin{align}
\label{eq:affine-support-fourier-main}
\wh p_{\bs_0,V}(\bu)=(-1)^{\bu\cdot \bs_0}\mathbbm{1}[\bu\in V^\perp].
\end{align}
Thus labels in $V^\perp$ are exactly the fully biased characters, and their signs determine whether
the affine-support $\bs_0+V$ contains the zero syndrome. If the support contains the zero syndrome,
then the remaining task is to certify that the distribution on $V$ is uniform. 
For this, one needs the nonzero Walsh characters on $V$. Labels that differ by an element of
$V^\perp$ have the same restriction to $V$, so the quotient $\Ftwo^n/V^\perp$ parametrizes the
Walsh characters on $V$. Thus choosing one representative of every nonzero coset of
$\Ftwo^n/V^\perp$ means choosing labels whose restrictions to $V$ realize all nontrivial Walsh
characters on $V$.

\begin{proposition}
\label{prop:structured-noise}
Run Alg.~\ref{alg:adaptive} with $\varepsilon=0$ on an
$r$-dimensional affine-support syndrome state $p_\rho=p_{\bs_0,V}$, and let
$d=n-r$. If, by some round $t$, the queried label set $Q_t$ contains a basis of $V^\perp$ and, in
the case $\bs_0\in V$, also contains one representative of every nonzero coset of
$\Ftwo^n/V^\perp$, equivalently every nontrivial Walsh character restricted to $V$, then
Alg.~\ref{alg:adaptive} terminates by round $t$ and returns
\begin{align}
F(\rho,\psi)=p_\rho(\boldsymbol{0})=
\begin{cases}
2^{-r}, & \bs_0\in V,\\
0, & \bs_0\notin V.
\end{cases}
\end{align}
Thus the number of structurally relevant stabilizer labels is $d$ if $\bs_0\notin V$, and at most
$d+2^r-1$ if $\bs_0\in V$.
\end{proposition}

Prop.~\ref{prop:structured-noise} states the performance of Alg.~\ref{alg:adaptive} in terms of the
labels it has actually acquired. It is a sufficient stopping condition based on the acquired label
set $Q_t$; it does not assert that the witness elimination policy necessarily discovers these labels
within $d+2^r-1$ queried stabilizers for every affine-support instance. For fixed support dimension
$r$, the number of structurally relevant labels is only $O(n)$, far below the $2^n-1$ labels needed
for full coverage. This is the sense in which affine-support structure can remove the
exponential information barrier: the algorithm does not need all Walsh characters, but only the
characters that certify the affine support and, when necessary, its uniformity.

The affine line case also shows why changing gauges is essential. For any predetermined gauge $A$
and any $n\ge2$, there exist two affine line distributions with identical generator expectations in
that gauge but fidelities $1/2$ and $0$. Here is the construction. 
Let $\be_1:=(1,0,\cdots,0)\in\Ftwo^n$ be the first standard basis
vector of $\Ftwo^n$.
In the coordinates of the chosen gauge, take
\begin{align}
\frac12(\delta_{\boldsymbol{0}}+\delta_{\boldsymbol{1}}),
\qquad
\frac12(\delta_{\be_1}+\delta_{\be_1+\boldsymbol{1}}),
\end{align}
and pull them back through $A$. Every coordinate generator has expectation zero for both
distributions, while only the first distribution contains the zero syndrome. Thus a fixed gauge can
leave a constant ambiguity on affine lines, whereas the adaptive search can learn the relevant
annihilator directions.

\begin{figure*}[!t]
\centering
\includegraphics[width=\textwidth]{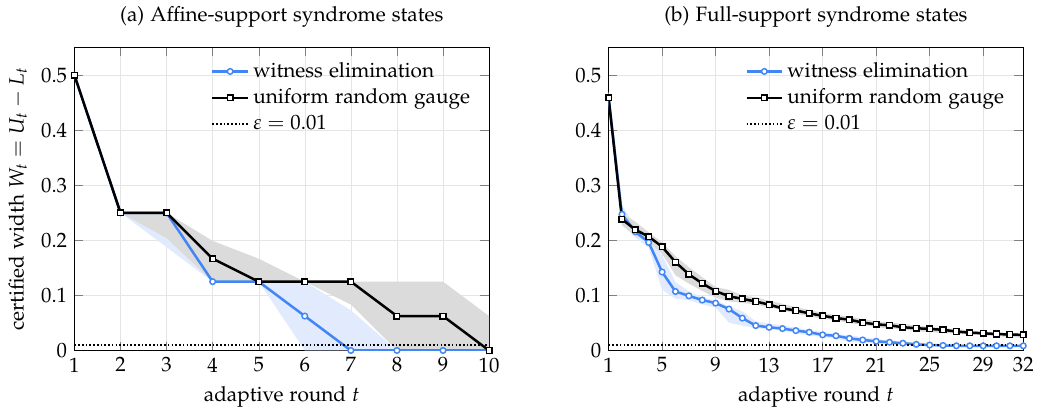}
\caption{\raggedright
\textbf{Witness elimination versus the uniform random gauge policy.}
Panel (a) uses $51$ affine-support syndrome states relative to $\ket{\mathrm{GHZ}_8}$, with
dimension $r=4$ and $F(\rho,\mathrm{GHZ}_8)=2^{-4}$. Panel (b) uses $51$ generic full-support
syndrome states relative to $\ket{\mathrm{GHZ}_8}$, sampled from the symmetric
Dirichlet distribution on $\Delta(\Ftwo^8)$. Curves show median certified width and
shaded bands show the interquartile range.}
\label{fig:numerics-witness-elimination}
\end{figure*}

\begin{table*}[t]
\centering
\setlength{\tabcolsep}{0pt}
\setlength\heavyrulewidth{0.3ex}
\renewcommand{\arraystretch}{1.5}
\begin{tabular}{@{}C{0.12\textwidth}C{0.3\textwidth}C{0.14\textwidth}C{0.12\textwidth}C{0.16\textwidth}C{0.12\textwidth}@{}}
\toprule
\multirow{2}{*}{\textbf{Ensemble}} & \multirow{2}{*}{\textbf{Settings}}
& \multicolumn{2}{c}{\textbf{Witness elimination}}
& \multicolumn{2}{c}{\textbf{Uniform random gauge}} \\
& & median $T_\varepsilon$ & failed runs & median $T_\varepsilon$ & failed runs \\
\midrule
affine-support & $n=8$, $r=4$, $51$ trials, $T_{\max}=10$
& $7$ & $1/51$ & $10$ & $25/51$ \\
full-support & $n=8$, $51$ trials, $T_{\max}=32$
& $25$ & $0/51$ & not reached & $51/51$ \\
\bottomrule
\end{tabular}
\caption{\raggedright
\textbf{Stopping statistics for the exact expectation benchmarks in Fig.~\ref{fig:numerics-witness-elimination}.}
Here $T_\varepsilon=\min\{t:W_t\le\varepsilon\}$ with $\varepsilon=0.01$ for both benchmarks.
A failed run is one with $W_t>\varepsilon$ at the cutoff $T_{\max}$. 
The phrase ``not reached'' means that the median run for
that policy did not reach the target width within the cutoff.}
\label{tab:numerics-stopping-rounds}
\end{table*}

\section{Numerical simulations}
\label{sec:numerical-simulations}

We conduct numerical simulations to support Alg.~\ref{alg:adaptive}. To keep the examples
consistent with the GHZ certificates in Sec.~\ref{sec:certified-upper-bound}, throughout this
section the target stabilizer state is the $n$-qubit GHZ state
\begin{align}
\ket{\mathrm{GHZ}_n}=\frac{\ket{0}^{\otimes n}+\ket{1}^{\otimes n}}{\sqrt{2}},
\end{align}
with reference stabilizer generators
\begin{align}\label{eq:GHZ-stabilizer-generators}
g_i=Z_iZ_{i+1} (i=1,\cdots,n-1);\;
g_n=X_1X_2\cdots X_n .
\end{align}
A syndrome $\bs\in\Ftwo^n$ records the signs $(-1)^{s_i}$ of these GHZ generators, so the zero
syndrome is the GHZ target sector and $F(\rho,\mathrm{GHZ}_n)=p_\rho(\boldsymbol{0})$. 

The simulations are performed directly in this syndrome representation. A quantum state instance $\rho$ 
is specified by a probability distribution $p_\rho$ on $\Ftwo^n$, the exact stabilizer expectation for
a queried label is $\mu_\rho(\bu)=\wh p_\rho(\bu)$, and each certified interval is obtained by
solving the two linear programs in Eqs.~\eqref{eq:certified-interval}
and~\eqref{eq:endpoint-witnesses}. This exact LP implementation has $2^n$ variables, so the
numerics below are intended to support the mechanism of the method on small and moderate system
sizes rather than to claim large-scale classical scalability.

\begin{figure*}[t]
\centering
\includegraphics[width=\textwidth]{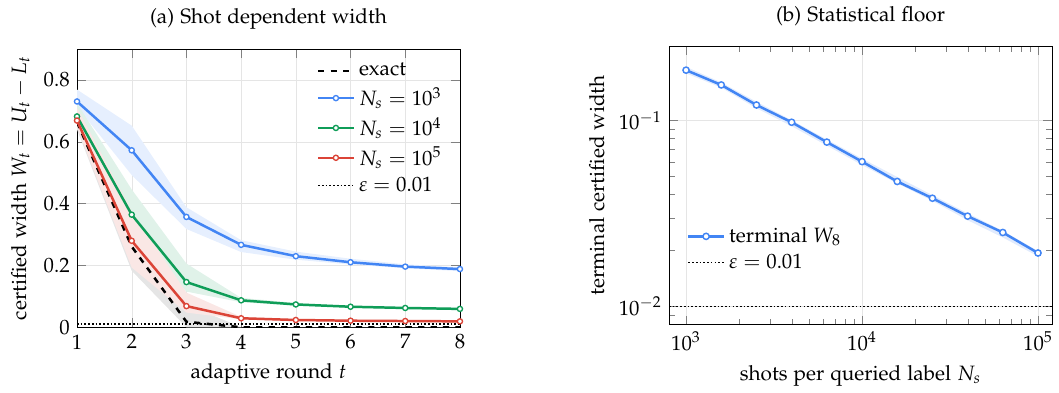}
\caption{\raggedright
\textbf{Effect of finite-shot noise.}
Each newly queried stabilizer expectation is estimated from $N_s$ independent $\pm1$ outcomes, and
every measured label is inserted into the finite-shot LP through the confidence interval constraint
$\widehat\mu(\bu)-\eta\le \wh p(\bu)\le\widehat\mu(\bu)+\eta$. Panel (a): median certified width over
$51$ trials for exact expectations and three representative shot budgets; shaded bands show the
interquartile range. Panel (b): median terminal width after the same eight adaptive rounds as
Panel (a), shown as a function of $N_s$ with the same interquartile summary. For a fixed $N_s$, the interval reaches a
statistical floor, and the floor decreases with $N_s$ as the shot budget increases. Across the
plotted shot budgets, the terminal interval contains the true fidelity in at least $50$ of the $51$
simulated trials.}
\label{fig:numerics-finite-shots}
\end{figure*}

\subsection{Witness elimination versus the uniform random gauge policy}\label{sec:numerical-witness-elimination}

The first set of experiments tests whether the witness elimination policy leads to faster
concentration than the uniform random gauge policy. To make the two benchmarks size matched, both
use $n=8$. In both cases, the target width is $\varepsilon=0.01$ and the number of random
instances is chosen to be approximately $20\%$ of the syndrome-space size $2^n$. 

We begin with the affine-support syndrome state from
Prop.~\ref{prop:structured-noise} because it is the structured case where the theory predicts a
small set of relevant labels: for each trial, $p_\rho$ is uniform on a random linear subspace
$V\le\Ftwo^n$ of dimension $r=4$, with $\bs_0=\boldsymbol{0}$. The true fidelity is
therefore $F(\rho,\mathrm{GHZ}_n)=p_\rho(\boldsymbol{0})=2^{-4}$. All runs start from the same
reference GHZ gauge. After the first round, we compare the witness elimination policy in
Alg.~\ref{alg:adaptive}
with the uniform random gauge policy over $51$ random instances. The cutoff is $T_{\max}=10$
rounds.
The left panel of Fig.~\ref{fig:numerics-witness-elimination} shows the result over $51$ random
instances. The uniform random gauge policy reduces the interval only gradually, while the witness
elimination policy typically collapses the remaining ambiguity several rounds earlier.

The affine-support ensemble is intentionally favorable to adaptive label discovery, 
so we also test a generic full-support ensemble without affine structure. 
In this second benchmark, $p_\rho$ is sampled from the symmetric Dirichlet distribution on the full simplex
$\Delta(\Ftwo^8)$ with all concentration parameters equal to one, i.e., every syndrome has nonzero probability almost surely. We use $51$ random
instances and $T_{\max}=32$. The right panel of
Fig.~\ref{fig:numerics-witness-elimination} shows that the witness elimination policy still
concentrates faster. The median final width is approximately $0.0085$ for witness elimination and
$0.028$ for the uniform random gauge policy. Thus the improvement is not limited to the
affine-support ensemble; the same endpoint disagreement criterion also gives faster concentration
for generic random syndrome distributions in this exact LP regime. The stopping statistics for both
benchmarks are summarized in Tab.~\ref{tab:numerics-stopping-rounds}.

\subsection{Finite shot effects}

The second experiment replaces exact access to $\mu_\rho(\bu)$ by a finite-shot adaptive variant of
Alg.~\ref{alg:adaptive}. The main algorithm uses exact equality constraints. In the finite-shot
variant, each newly queried stabilizer label $\bu$ is measured with $N_s$ shots, while labels already
measured in earlier rounds retain their existing confidence intervals. We simulate independent
outcomes $X_j\in\{\pm1\}$ with mean $\mu_\rho(\bu)$ and use
\begin{align}
\widehat\mu(\bu)=\frac{1}{N_s}\sum_{j=1}^{N_s}X_j .
\end{align}
To retain a certified feasible set, every measured label enters the endpoint LP through an interval
constraint, not through an equality constraint. Specifically, after round $t$ we use the relaxed
finite-shot feasible set
\begin{align}
\label{eq:finite-shot-feasible-polytope}
\cF_t^{\mathrm{fs}}
&:= \{p\in\Delta(\Ftwo^n): \nonumber\\
&\qquad\;\; \widehat\mu(\bu)-\eta
\le \wh p(\bu) \le \widehat\mu(\bu)+\eta
\; \forall \bu\in Q_t\},
\end{align}
with each interval clipped to the physical range $[-1,1]$, and define the reported finite-shot
endpoints by minimizing and maximizing $p(\boldsymbol{0})$ over $\cF_t^{\mathrm{fs}}$. We choose
\begin{align}
\eta&=\sqrt{\frac{2\log(2nT_{\max}/\delta)}{N_s}},
\end{align}
which is the Hoeffding radius for $\pm1$ outcomes with a union bound over at most $nT_{\max}$ newly
queried labels.

The adaptive choice of labels does not invalidate this confidence statement. Let $\cH_{t-1}$ be
the finite-shot history before round $t$. Conditional on $\cH_{t-1}$, the gauge selected in round
$t$, and hence the newly measured labels, are fixed by the policy. The shot outcomes for those
labels are independent with means $\mu_\rho(\bu)$, so Hoeffding's inequality applies conditionally
to each selected label. A union bound over all selected labels in all rounds up to $T_{\max}$ gives,
with probability at least $1-\delta$, simultaneous containment of every true Walsh coefficient in
its confidence interval. On this event, the true syndrome distribution $p_\rho$ belongs to
$\cF_t^{\mathrm{fs}}$ for every $t\le T_{\max}$, so every reported finite-shot interval contains the
true fidelity $F(\rho,\psi)$.

Fig.~\ref{fig:numerics-finite-shots} uses sparse syndrome distributions relative to the
$8$-qubit GHZ target, with fidelity $F(\rho,\mathrm{GHZ}_8)=0.64$ and five nonzero error
syndromes. As in Sec.~\ref{sec:numerical-witness-elimination}, the target width is fixed to
$\varepsilon=0.01$ and the number of random instances is $51$, approximately $20\%$ of
the syndrome-space size $2^8$. Panel (a) compares the exact adaptive curve with
representative finite-shot runs at $N_s=10^3,10^4,10^5$, while Panel (b) uses eleven
log-spaced shot budgets from $10^3$ to $10^5$. We use $\delta=0.05$ and run every curve for
eight adaptive rounds. Exact
expectations allow the interval to collapse in a few rounds. With a fixed finite-shot budget, the
interval contracts only down to a statistical floor set by $\eta$. This is not a failure of
adaptivity: it is the expected price of retaining a certified confidence region. The floor decreases
as the shot budget increases. The median terminal widths after eight rounds are approximately $0.186$, $0.060$,
and $0.019$ for the three representative shot budgets, respectively. At the largest shot budget, 
the finite-shot floor is close to, but still above, the target width $\varepsilon=0.01$.
Across the plotted shot budgets, the true fidelity remains inside the terminal confidence interval in
at least $50$ of the $51$ trials.

\section{Comparison with existing works}\label{sec:comparison}
This work differs from existing methods in its measurement interface, inference target, and type of
guarantee. These distinctions show that the methods below solve certification tasks different from
ours.

\subsection{Comparison with direct fidelity estimation}
Direct fidelity estimation (DFE)~\cite{FlammiaLiu2011} starts from the following stabilizer identity:
\begin{align}
\label{eq:dfe-average}
F(\rho,\psi)=2^{-n}\sum_{g\in S}\tr\left[\rho g\right]
=\bE_{g\sim \mathrm{Unif}(S)}\,\tr\left[\rho g\right],
\end{align}
where $\mathrm{Unif}(S)$ denotes the uniform distribution on $S$.
DFE treats Eq.~\eqref{eq:dfe-average} as a Monte Carlo average over arbitrary stabilizer elements.
It estimates the fidelity additively by sampling
$O(\varepsilon^{-2}\log(1/\delta))$ stabilizers from $S$, with constant shot cost per sampled
observable for stabilizer targets. 
This is why the DFE observable count does not scale with the number of qubits $n$.

Our protocol solves a different task. It retains only the generator expectations in the queried
gauges and outputs a certified interval over all syndrome distributions compatible with those data.
The distinction can be seen directly. Suppose the queried labels after round $t$ are
$Q_t\subseteq\Ftwo^n\setminus\{\boldsymbol{0}\}$ and some nontrivial label
$\bv\notin Q_t$ remains unqueried. Define two syndrome distributions
\begin{align}
p_\pm(\bs)=2^{-n}\left(1\pm(-1)^{\bv\cdot\bs}\right).
\end{align}
Both distributions are valid. Their Walsh coefficients agree on every queried label in $Q_t$, but
their fidelities differ by $p_+(\boldsymbol{0})-p_-(\boldsymbol{0})=2^{1-n}$.
Thus the same generator gauge data can be compatible with two fidelities separated by $2^{1-n}$.
Any valid monotone certified interval for this data record must therefore have width at least
$2^{1-n}$. Exact certification, or a uniform guarantee of width below $2^{1-n}$, requires all
$2^n-1$ nontrivial stabilizer labels and hence at least $\lceil(2^n-1)/n\rceil$ full gauge rounds.
This does not contradict DFE: DFE estimates an average by sampling arbitrary stabilizers, whereas
generator gauge certification asks for a worst-case compatibility guarantee under a weaker retained
data model.

\subsection{Comparison with KKL fixed-gauge certification}
Our closest antecedent is the KKL protocol~\cite{Kalev2019}. 
The measurement interface is the same: one measures
$n$ commuting generators in a chosen gauge and computes the worst-case lower fidelity consistent
with these expectations. 
Sec.~\ref{sec:certified-upper-bound} complements this result by giving the companion upper endpoint 
and its finite-shot certificate. 
To summarize, the KKL supplies the lower endpoint $L_1$, while
we provide the upper endpoint $U_1$ from the same retained data.

The adaptive stabilizer fidelity certification algorithm described in Sec.~\ref{sec:adaptive-framework} 
solves a different problem: KKL fixes the gauge, whereas we treat the gauge as a
design variable and then adapt it across rounds. Thus the one gauge interval is only the first step
of our certified interval framework.
Indeed, the examples elaborated in Sec.~\ref{sec:illustrative-examples} show the consequence. 
Two gauges for the same
state can produce very different lower certificates, and one additional gauge can collapse the
feasible region even when one non-identity stabilizer has not been queried. Our contribution is
therefore both a complete one gauge interval and a policy for choosing which new gauge should be
queried next.

\subsection{Comparison with stabilizer state learning}
Stabilizer states also admit efficient learning algorithms under learning-oriented access models.
For example, stabilizer states are PAC-learnable~\cite{Rocchetto2018}, and Bell sampling can
identify an unknown stabilizer state from $O(n)$ copies~\cite{Montanaro2017}. These results
build on the fact that stabilizer states have compact classical descriptions~\cite{AaronsonGottesman2004}.

These learning results solve a different problem from the one considered here. A learning algorithm
aims to output a hypothesis stabilizer description, or a predictor for future measurement outcomes,
under a specified learning model. Our target state is already known. The unknown object is the
prepared, possibly nonstabilizer, noisy state, and the output is not a learned stabilizer state but a
certified interval $[L_t,U_t]$ for its fidelity with the target. The guarantee is adversarial: every
state compatible with the retained generator gauge expectations must have fidelity inside the
reported interval. 
The measurement interface is also different: stabilizer learning uses learning oriented access such
as Bell sampling or PAC queries, whereas our protocol retains only generator expectations from
chosen gauges of the known target stabilizer group.

\subsection{Comparison with adaptive tomography}
Adaptive tomography with active learning \cite{Lange2023} aims to reconstruct an unknown quantum
state by choosing informative measurement configurations. Our method uses the same adaptive
philosophy but targets a different object: a certified interval for fidelity to a known stabilizer
state. The representation is the syndrome distribution rather than the full density operator, and
the allowed actions are generator gauges rather than general tomographic settings.

Briefly, adaptive tomography seeks global reconstruction, 
whereas our framework seeks the tightest certified fidelity interval available from generator gauge data. 
The shared lesson is that adaptive measurement choice can help, but the tasks and guarantees are different.

\section{Conclusions and outlook}\label{sec:conclusions}
We have developed an adaptive, interval-valued framework for stabilizer fidelity certification from
generator gauge data. Starting from the KKL fixed-gauge lower certificate, we first derived the
complementary optimal worst-case upper certificate for one chosen gauge, thereby completing the
single-gauge fidelity interval. We then treated the generator gauge as an adaptive design variable:
each round updates the feasible syndrome polytope, reports the certified interval $[L_t,U_t]$, and
chooses the next gauge by the witness elimination policy. This formulation makes the gauge
dependence of fixed-gauge certification explicit and turns additional gauges into new Walsh
constraints on the syndrome distribution.

Our analysis establishes both guarantees and limitations. The certified interval tightens
monotonically and collapses to the true fidelity once all nontrivial stabilizer labels have been
covered, while the worst-case lower bound shows that arbitrary states may still require full
coverage. Randomized gauge policies give conservative coverage baselines, and affine-support
syndrome states provide a structured regime in which the exponential full-coverage benchmark can
be avoided. The numerical simulations support these conclusions: witness elimination concentrates
faster than the uniform random gauge policy for both affine-support and generic full-support
ensembles, and the finite-shot study shows the expected statistical floor as the shot budget varies.
We also clarified how our results differ from existing direct fidelity estimation, 
KKL fixed-gauge certification, stabilizer state learning, and adaptive tomography.

Two directions remain important. First, the exact LP over the full syndrome simplex
should be replaced or complemented by scalable relaxations that preserve useful certificate
guarantees. Second, a more complete finite-shot theory should integrate noisy stabilizer estimates,
adaptive gauge choice, and statistically valid confidence intervals in a single end-to-end
guarantee.

\section*{Acknowledgements}

This work was supported by 
the Quantum Science and Technology-National Science and Technology Major Project (Grant No. 2025ZD0300300) and 
the National Natural Science Foundation of China (Grant No. 12504584).

%

\makeatletter
\newcommand{\appendixtitle}[1]{\gdef\@title{#1}}
\makeatother

\makeatletter%
\newcommand{\appendixmaketitle}{%
\begin{center}%
\vspace{0.4in}%
{\large \@title \par}%
\end{center}%
\par%
}%
\makeatother%

\makeatletter
\newcommand{\appendixtableofcontents}{%
\begingroup
\setcounter{tocdepth}{4}%
\@starttoc{atoc}%
\endgroup}
\makeatother

\newcommand{\appendixtocdivider}{%
\par\medskip
\noindent\hbox to \linewidth{%
\leaders\hbox{\rule[0.6ex]{1pt}{0.4pt}}\hfill
\hspace{0.8em}\textsc{Table of Contents}\hspace{0.8em}%
\leaders\hbox{\rule[0.6ex]{1pt}{0.4pt}}\hfill}%
\par}

\makeatletter
\newcounter{subsubsubsection}[subsubsection]
\renewcommand{\thesubsubsubsection}{\thesubsubsection.\arabic{subsubsubsection}}
\providecommand{\theHsubsubsubsection}{}
\renewcommand{\theHsubsubsubsection}{\theHsubsubsection.\arabic{subsubsubsection}}
\providecommand*{\toclevel@subsubsubsection}{4}
\newcommand{\subsubsubsection}[1]{%
\refstepcounter{subsubsubsection}%
\paragraph*{\thesubsubsubsection\space #1}}
\newcommand*\l@subsubsubsection[2]{\@dottedtocline{4}{4.4em}{3.0em}{#1}{#2}}
\makeatother

\newcommand{\appsection}[1]{%
\section{#1}%
\addcontentsline{atoc}{section}{\protect\numberline{\thesection}#1}}

\newcommand{\appsubsection}[1]{%
\subsection{#1}%
\addcontentsline{atoc}{subsection}{\protect\numberline{\thesubsection}#1}}

\newcommand{\appsubsubsection}[1]{%
\subsubsection{#1}%
\addcontentsline{atoc}{subsubsection}{\protect\numberline{\thesubsubsection}#1}}

\newcommand{\appsubsubsubsection}[1]{%
\subsubsubsection{#1}}

\setcounter{secnumdepth}{4}
\appendix
\widetext
\newpage

\appendixtitle{\bf 
Supplemental Material for\\``\thetitle''}
\appendixmaketitle
\vspace{0.1in}

This Supplemental Material is organized to follow the structure of the main text (MT). 
Appx.~\ref{app:drawbacks-KKL-certificate} gives the details for Sec.~\ref{sec:drawbacks-KKL-certificate}; 
Appx.~\ref{app:certified-upper-bound} proves the one gauge upper endpoint results in Sec.~\ref{sec:certified-upper-bound}; 
Appx.~\ref{app:fine-grained-certification-algorithm} describes the single stabilizer refined algorithm 
associated with Sec.~\ref{sec:adaptive-framework}; 
and Appx.~\ref{app:performance-guarantees} proves the performance guarantees in Sec.~\ref{sec:performance-guarantees}.

\appendixtocdivider
{%
\appendixtableofcontents%
}%

\appsection{Additional material for Sec.~\ref{sec:drawbacks-KKL-certificate}}
\label{app:drawbacks-KKL-certificate}

\appsubsection{Proof of Prop.~\ref{prop:singlebasis-insufficient}}
\label{app:singlebasis-insufficient}
Let $A\in\GL(n,2)$ be arbitrary. In the $A$-coordinates, define two distributions on $\Ftwo^n$ by
\begin{align}
\wt p=\frac12\bigl(\delta_{\boldsymbol{0}}+\delta_{\be_1+\be_2}\bigr),
\qquad
\wt q=\frac12\bigl(\delta_{\be_1}+\delta_{\be_2}\bigr),
\end{align}
where $\be_1,\be_2$ are the first two standard basis vectors. For every coordinate vector $\be_i$,
\begin{align}
\wh{\wt p}(\be_i)=\wh{\wt q}(\be_i)=
\begin{cases}
0, & i=1,2,\\
1, & i=3,\cdots,n,
\end{cases}
\end{align}
because the two support points in each distribution have opposite parity in the $i$th coordinate whenever $i\in\{1,2\}$, and identical zero parity otherwise. However,
\begin{align}
\wt p(\boldsymbol{0})=\frac12,
\qquad
\wt q(\boldsymbol{0})=0.
\end{align}
Pull these distributions back to the reference coordinates by defining
\begin{align}
p(\bs):=\wt p(A^{\mathsf T}\bs),\qquad q(\bs):=\wt q(A^{\mathsf T}\bs).
\end{align}
Since the columns of $A$ are $\ba_1,\cdots,\ba_n$, the degree-one characters in the $A$-coordinates are precisely the queried characters $\wh p(\ba_i)$ and $\wh q(\ba_i)$. Hence
\begin{align}
\wh p(\ba_i)=\wh q(\ba_i)\qquad \forall i\in[n],
\end{align}
while
\begin{align}
p(\boldsymbol{0})=\wt p(\boldsymbol{0})=\frac12\neq 0=\wt q(\boldsymbol{0})=q(\boldsymbol{0}).
\end{align}
This proves the proposition.

\appsubsection{Further details for the illustrative examples}
\label{app:examples}
For the gauge choice claim in Sec.~\ref{sec:illustrative-examples}, the KKL lower bound from one gauge for generator expectations $\mu_1,\cdots,\mu_n$ is
\begin{align}
L_{\mathrm{KKL}}=\max\left\{0,1-\frac12\sum_{i=1}^n(1-\mu_i)\right\},
\end{align}
which is the explicit worst-case fidelity formula derived in Ref.~\cite{Kalev2019}. Substituting $(\mu_1,\mu_2,\mu_3)=(1/2,1/2,1/2)$ gives $L_{\mathrm{KKL}}=1/4$, while $(\mu_1,\mu_2,\mu_3)=(0,0,-1/2)$ gives $L_{\mathrm{KKL}}=0$.

For the additional gauge claim in Sec.~\ref{sec:illustrative-examples}, the parity gauge gives
\begin{align}
\wh p(\boldsymbol{110})=0,\qquad \wh p(\boldsymbol{101})=0,\qquad \wh p(\boldsymbol{111})=-1/2.
\end{align}
The added canonical gauge gives
\begin{align}
\wh p(\boldsymbol{100})=\wh p(\boldsymbol{010})=\wh p(\boldsymbol{001})=1/2.
\end{align}
The queried labels are therefore $\boldsymbol{100},\boldsymbol{010},\boldsymbol{001},\boldsymbol{110},\boldsymbol{101},\boldsymbol{111}$, so $\boldsymbol{011}$ is not queried. Let
\begin{align}
x:=\wh p(\boldsymbol{011})
\end{align}
be the only unqueried Walsh coefficient. Walsh inversion gives
\begin{align}
p_{000}&=\frac14+\frac{x}{8},&
p_{100}&=\frac14+\frac{x}{8},\nonumber\\
p_{010}&=\frac14-\frac{x}{8},&
p_{001}&=\frac14-\frac{x}{8},\\
p_{011}&=\frac{x}{8},&
p_{111}&=\frac{x}{8},\nonumber\\
p_{101}&=-\frac{x}{8},&
p_{110}&=-\frac{x}{8}.\nonumber
\end{align}
Positivity of $p_{011}$ and $p_{111}$ implies
\begin{align}
x\ge 0,
\end{align}
whereas positivity of $p_{101}$ and $p_{110}$ implies
\begin{align}
x\le 0.
\end{align}
Hence $x=0$, so the distribution is forced to be uniform on $000,100,010,001$. Thus the certified fidelity interval is exact, with $p_{000}=1/4$, even though the stabilizer $Z_2Z_3$ was never queried.

\appsection{Proofs for Sec.~\ref{sec:certified-upper-bound}}
\label{app:certified-upper-bound}
\appsubsection{Proof of Prop.~\ref{prop:single-gauge-upper}}
\label{app:single-gauge-upper}
Work in the coordinates of the chosen gauge $A$, so that the measured generator labels are the
standard basis vectors $\be_1,\cdots,\be_n$. Write
\begin{align}
a_i:=\frac{1+\mu_\rho(\ba_i)}{2}.
\end{align}
For any feasible syndrome distribution $p$, the constraint on $\ba_i$ is equivalent to
\begin{align}
\sum_{\bs:s_i=0}p(\bs)=a_i.
\end{align}
The event $\{\bs=\boldsymbol{0}\}$ is contained in every event $\{s_i=0\}$, so every feasible
distribution satisfies
\begin{align}
p(\boldsymbol{0})\le a_i\qquad \forall i\in[n].
\end{align}
Therefore
\begin{align}
p(\boldsymbol{0})\le \min_{i\in[n]}a_i.
\end{align}

It remains to show that this upper bound is attainable. After a permutation of the chosen gauge
columns, assume that
\begin{align}
a_1\le a_2\le \cdots \le a_n.
\end{align}
Define a probability distribution in the chosen gauge coordinates by placing mass on the nested
syndromes
\begin{align}
\boldsymbol{0},\quad
\be_1,\quad
\be_1+\be_2,\quad \cdots,\quad
\be_1+\cdots+\be_n
\end{align}
as follows:
\begin{align}
p(\boldsymbol{0})&=a_1,\nonumber\\
p(\be_1+\cdots+\be_j)&=a_{j+1}-a_j,\qquad j=1,\cdots,n-1,\nonumber\\
p(\be_1+\cdots+\be_n)&=1-a_n,
\end{align}
and set all other syndrome probabilities to zero. These probabilities are nonnegative and sum to
one. For coordinate $i$, the syndromes with $s_i=0$ are exactly the first $i$ support points in this
nested list, so their total probability is $a_i$. Hence the distribution satisfies all one gauge
constraints and has
\begin{align}
p(\boldsymbol{0})=a_1=\min_{i\in[n]}a_i.
\end{align}
Transforming back from the chosen gauge coordinates to the reference coordinates preserves the value
of $p(\boldsymbol{0})$ and the measured generator expectations. Thus the upper bound is tight, which
proves Eq.~\eqref{eq:kkl-upper-bound}.

The analytic formulas also show directly that the one gauge fidelity interval is indeed an
interval. Notice that Eqs.~\eqref{eq:kkl-bound} and~\eqref{eq:kkl-upper-bound} can be equivalently written as
\begin{align}
F_{\min}(A;\rho)&=\max\left\{0,\,1-\sum_{i=1}^n(1-a_i)\right\},\\
F_{\max}(A;\rho)&=\min_{i\in[n]}a_i.
\end{align}
Since $a_i\ge0$, if $1-\sum_i(1-a_i)\le0$, then $F_{\min}(A;\rho)=0\le F_{\max}(A;\rho)$. If
$1-\sum_i(1-a_i)>0$, then for every $i\in[n]$,
\begin{align}
1-\sum_{j=1}^n(1-a_j)
=a_i-\sum_{j\neq i}(1-a_j)
\le a_i,
\end{align}
because each $a_j\le1$. Taking the minimum over $i$ gives
\begin{align}
F_{\min}(A;\rho)
\le
F_{\max}(A;\rho).
\end{align}

\appsubsection{Proof of Prop.~\ref{prop:single-gauge-upper-finite}}
\label{app:single-gauge-upper-finite}
For each $i$, Hoeffding's inequality for the binary variables $Y_{i,j}\in\{0,1\}$ gives
\begin{align}
\Pr\left\{|\widehat a_i-a_i|>\varepsilon\right\}
\le 2\exp(-2m_i\varepsilon^2).
\end{align}
Under Eq.~\eqref{eq:upper-shot-per-generator}, the right-hand side is at most $\delta/n$.
A union bound over $i\in[n]$ gives
\begin{align}
\max_{i\in[n]}|\widehat a_i-a_i|\le \varepsilon
\end{align}
with probability at least $1-\delta$. On this event,
\begin{align}
\left|\min_{i\in[n]}\widehat a_i-\min_{i\in[n]}a_i\right|
\le \max_{i\in[n]}|\widehat a_i-a_i|
\le \varepsilon .
\end{align}
By Prop.~\ref{prop:single-gauge-upper}, $\min_i a_i=F_{\max}(A;\rho)$, which proves
Eq.~\eqref{eq:upper-finite-confidence}.

\appsubsection{Proof of Cor.~\ref{cor:single-gauge-upper-sample}}
\label{app:single-gauge-upper-sample}
The true syndrome distribution of $\rho$ is feasible for the exact optimization problem in
Eq.~\eqref{eq:kkl-upper-polytope-form}. Hence
\begin{align}
F(\rho,\psi)\le F_{\max}(A;\rho).
\end{align}
Combining this inequality with Prop.~\ref{prop:single-gauge-upper-finite} gives
Eq.~\eqref{eq:finite-upper-certificate} with probability at least $1-\delta$. Summing the
per generator shot count in Eq.~\eqref{eq:upper-shot-per-generator} over all $n$ generators gives
Eq.~\eqref{eq:upper-total-shots}.

\appsubsection{Calculation of the GHZ upper certificates}
\label{app:ghz-upper-calculation}
Here we spell out the upper certificate entries in Tab.~\ref{tab:kkl-ghz-upper}. For a desired
confidence level $p_{\mathrm{conf}}$, set $\delta=1-p_{\mathrm{conf}}$. Cor.~\ref{cor:single-gauge-upper-sample}
gives
\begin{align}
F(\rho,\psi)\le \widehat U_A+\varepsilon,
\qquad
\varepsilon=
\sqrt{\frac{\log(2n/\delta)}{2m}},
\end{align}
when each of the $n$ generators has effective sample count $m$. Since every fidelity is at most
$1$, the reported upper certificate is
\begin{align}
U_{\mathrm{cert}}
=\min\{1,\widehat U_A+\varepsilon\}.
\end{align}
This last clipping step is why the trapped-ion upper entries equal $1.000$ even though
$\widehat U_A=0.979$.

For the trapped-ion data, $n=3$, $m=11000/3$, and $\widehat U_A=0.979$. The three confidence levels
give
\begin{align}
\varepsilon_{0.999}=0.034,\qquad
\varepsilon_{0.990}=0.030,\qquad
\varepsilon_{0.900}=0.024.
\end{align}
Thus $\widehat U_A+\varepsilon$ is respectively $1.013$, $1.009$, and $1.003$, and all three values
are reported as $1.000$ after applying the physical upper bound. For the IBM data,
$m=8192/3$ and $\widehat U_A=0.838$, giving
\begin{align}
0.838+0.040=0.878,\qquad
0.838+0.034=0.872,\qquad
0.838+0.027=0.865,
\end{align}
which are the IBM upper certificate entries in Tab.~\ref{tab:kkl-ghz-upper}.

\appsection{Fine grained adaptive stabilizer fidelity certification algorithm}
\label{app:fine-grained-certification-algorithm}

Fig.~\ref{fig:hyperplane-cut} suggests a finer acquisition policy than the gauge-level protocol in
the main text. The disagreement score $d_t(\bu)$ is defined for each individual stabilizer label
$\bu$, whereas Alg.~\ref{alg:adaptive} must package its choices into a full generator gauge. This
packaging is natural for the generator-only KKL interface, but it can be restrictive if the
experimental interface allows an arbitrary stabilizer $g(\bu)$ to be queried as a standalone
observable. In this stronger, fine grained access model, one can query the single most informative
stabilizer at each step.

\begin{algorithm}[t]
\captionsetup{justification=raggedright,singlelinecheck=false}
\caption{Fine grained adaptive stabilizer fidelity certification}
\label{alg:single-stabilizer}
\noindent%
\begin{tabular}{@{}>{\raggedright\arraybackslash}p{4.8em}@{\hspace{0.35em}}>{\raggedright\arraybackslash}p{\dimexpr\linewidth-5.15em\relax}@{}}
\textbf{Input:} & Target stabilizer state $\ket{\psi}$; unknown state $\rho$; threshold $\varepsilon\ge 0$.\\
\textbf{Output:} & A sequence of certified intervals $[L_t,U_t]$ for $F(\rho,\psi)$.
\end{tabular}%
\par\vspace{-4pt}
\begin{algorithmic}[1]
\STATE Fix reference generators $g_1,\cdots,g_n$ for $\ket{\psi}$.
\STATE Choose an initial gauge $A_0\in\GL(n,2)$ with columns $\ba_{0,1},\cdots,\ba_{0,n}$.
\STATE \textbf{Initial generator query:} measure $g(\ba_{0,i})$ for all $i\in[n]$ and obtain $\mu_\rho(\ba_{0,i})$.
\STATE Set $Q_0\gets\{\ba_{0,1},\cdots,\ba_{0,n}\}$.
\FOR{$t=0,1,2,\cdots$}
    \STATE Compute the feasible polytope $\cF_t$ via Eq.~\eqref{eq:single-feasible-polytope}.
    \STATE Compute the two endpoint witnesses as in Eq.~\eqref{eq:endpoint-witnesses}.
    \STATE Set $L_t\gets p_t^-(\boldsymbol{0})$, $U_t\gets p_t^+(\boldsymbol{0})$, and $W_t\gets U_t-L_t$.
    \STATE Record the certified interval $[L_t,U_t]$.
    \IF{$W_t\leq\varepsilon$ or $Q_t=\Ftwo^n\setminus\{\boldsymbol{0}\}$}
        \RETURN All recorded certified intervals.
    \ENDIF
    \STATE Compute $d_t(\bu)\gets|\wh p_t^+(\bu)-\wh p_t^-(\bu)|$ for every unqueried nonzero label $\bu$.
    \STATE Choose $\bu_{t+1}$ by Eq.~\eqref{eq:next-single-stabilizer}.
    \STATE \textbf{Single-label query:} measure only $g(\bu_{t+1})$ and obtain $\mu_\rho(\bu_{t+1})$.
    \STATE Set $Q_{t+1}\gets Q_t\cup\{\bu_{t+1}\}$.
\ENDFOR
\end{algorithmic}
\end{algorithm}

\vspace*{0.1in}
\textbf{Initial generator query.}
The fine grained certification algorithm still starts from a KKL gauge. To make the comparison with
Alg.~\ref{alg:adaptive} direct, it uses the same input data rather than assuming that a single-gauge
certificate has already been measured. It first fixes reference generators for $\ket{\psi}$, chooses
an initial gauge $A_0$ with columns $\ba_{0,1},\cdots,\ba_{0,n}$, and queries the corresponding
$n$ stabilizer expectations. After this common initialization, set
\begin{align}
Q_0:=\{\ba_{0,1},\cdots,\ba_{0,n}\}.
\end{align}
This initialization is necessary. Suppose only a queried label set
$R\subseteq\Ftwo^n\setminus\{\boldsymbol{0}\}$ with $\operatorname{rank}(R)<n$ is available. One can
still write partial-data lower and upper LPs by optimizing $p(\boldsymbol{0})$ over all syndrome
distributions satisfying $\wh p(\bu)=\mu_\rho(\bu)$ for $\bu\in R$, in the same spirit as
Eqs.~\eqref{eq:kkl-polytope-form} and~\eqref{eq:kkl-upper-polytope-form}. However, these
partial-data endpoints can be completely vacuous. Since $\operatorname{rank}(R)<n$, there exists a
nonzero syndrome $\bs_\star\in\Ftwo^n$ such that $\bu\cdot\bs_\star=0$ for every $\bu\in R$. The
two syndrome distributions
\begin{align}
p=\delta_{\boldsymbol{0}},
\qquad
q=\delta_{\bs_\star}.
\end{align}
then satisfy
\begin{align}
\wh p(\bu)=1=\wh q(\bu),\qquad \forall \bu\in R,
\end{align}
but have fidelities $p(\boldsymbol{0})=1$ and $q(\boldsymbol{0})=0$. Hence the partial-data interval
is $[0,1]$ for this valid data set. Fewer than $n$ independent labels therefore cannot, in general,
give nontrivial certified lower and upper bounds.

The closed one gauge formulas in Eqs.~\eqref{eq:kkl-bound} and~\eqref{eq:kkl-upper-bound} are
therefore certificates only after a full generator gauge has been queried. Applying them earlier
would either require unmeasured expectations from a completion of $R$ to a basis, or would
incorrectly treat the incomplete set as a gauge. Once $A_0\in\GL(n,2)$ has been queried, its columns
form a basis of $\Ftwo^n$ and the measured constraints are precisely
\begin{align}
\wh p(\ba_{0,i})=\mu_\rho(\ba_{0,i}),\qquad i\in[n].
\end{align}
Thus the resulting feasible set is exactly the one-gauge feasible polytope in
Eqs.~\eqref{eq:kkl-polytope-form} and~\eqref{eq:kkl-upper-polytope-form} with $A=A_0$. The initial
generator query is therefore not an additional resource beyond the gauge-level protocol; it is the
common KKL starting point needed for a fair comparison.

\vspace*{0.1in}
\textbf{Single-label query.}
After the initial generator query, the refined adaptive procedure is to choose one stabilizer, 
rather than a gauge, at a time. 
This is the loop implemented in Alg.~\ref{alg:single-stabilizer}: at each nonterminal step, the algorithm recomputes
the endpoint witnesses, queries one unmeasured stabilizer, and updates the cumulative queried set.
For any current queried set $Q_k\subseteq\Ftwo^n\setminus\{\boldsymbol{0}\}$, define
\begin{align}
\label{eq:single-feasible-polytope}
\cF_k:=\{p\in\Delta(\Ftwo^n):\wh p(\bu)=\mu_\rho(\bu)\ \forall \bu\in Q_k\},
\end{align}
and define $L_k,U_k,W_k,p_k^-$, and $p_k^+$ exactly as in
Eqs.~\eqref{eq:certified-interval} and~\eqref{eq:endpoint-witnesses}. The disagreement score is
again
\begin{align}
\label{eq:single-disagreement-score}
d_k(\bu):=\left|\wh p_k^+(\bu)-\wh p_k^-(\bu)\right|,
\qquad \bu\in\Ftwo^n\setminus\{\boldsymbol{0}\}.
\end{align}
The next stabilizer is chosen by the greedy single-label policy
\begin{align}
\label{eq:next-single-stabilizer}
\bu_{k+1}\in
\argmax_{\bu\in(\Ftwo^n\setminus\{\boldsymbol{0}\})\setminus Q_k}d_k(\bu).
\end{align}
After querying $g(\bu_{k+1})$ and obtaining $\mu_\rho(\bu_{k+1})$, update
\begin{align}
Q_{k+1}:=Q_k\cup\{\bu_{k+1}\}.
\end{align}

\vspace*{0.1in}
\textbf{Comparison with Alg.~\ref{alg:adaptive}.}
Both algorithms have the same input and output, and both first acquire $n$ independent stabilizer
expectations through an initial gauge. If the same initial gauge is used, their first certified
interval is the same one-gauge KKL interval. The difference begins after this initialization:
\begin{itemize}
\item Alg.~\ref{alg:adaptive} continues to acquire data in full gauges, adding up to $n$
      independent labels before the next endpoint-LP solve.
\item Alg.~\ref{alg:single-stabilizer} adds only the single unqueried label with largest endpoint
      disagreement and then resolves the endpoint LPs.
\end{itemize}
Alg.~\ref{alg:single-stabilizer} avoids completing a gauge with low-disagreement labels, but it uses a
different single stabilizer measurement interface and can require about $n$ endpoint optimization
cycles to acquire $n$ additional labels.

\vspace*{0.1in}
\textbf{Performance analysis for Alg.~\ref{alg:single-stabilizer}.}
The performance analysis for Alg.~\ref{alg:single-stabilizer} follows the same logic as the analysis
of Alg.~\ref{alg:adaptive}. Because each single-label query adds one exact Walsh constraint and never
removes previous constraints, the certified interval is monotone. Once all nontrivial stabilizer
labels have been queried, Walsh--Hadamard inversion identifies the syndrome distribution exactly,
so the interval collapses to the true fidelity. Conversely, the same worst-case obstruction shows
that full coverage can be necessary in the absence of additional structure.

\appsection{Proofs for Sec.~\ref{sec:performance-guarantees}}
\label{app:performance-guarantees}

This appendix proves the performance guarantees in Sec.~\ref{sec:performance-guarantees} in the
same order in which they appear in the main text: monotonicity, full-coverage completeness, the
worst-case lower bound, randomized coverage, endpoint disagreement, and the affine-support
structured-state guarantee.

\appsubsection{Proof of Prop.~\ref{prop:monotonicity}}
\label{app:monotonicity}
Since the constraints are cumulative, $Q_t\subseteq Q_{t+1}$. Hence every distribution satisfying all constraints at round $t+1$ also satisfies all constraints at round $t$, so
\begin{align}
\cF_{t+1}\subseteq \cF_t.
\end{align}
Therefore
\begin{align}
L_{t+1}=\min_{p\in\cF_{t+1}}p(\boldsymbol{0})\ge \min_{p\in\cF_t}p(\boldsymbol{0})=L_t,
\end{align}
and similarly
\begin{align}
U_{t+1}=\max_{p\in\cF_{t+1}}p(\boldsymbol{0})\le \max_{p\in\cF_t}p(\boldsymbol{0})=U_t.
\end{align}
It follows immediately that $W_{t+1}=U_{t+1}-L_{t+1}\le U_t-L_t=W_t$.

\appsubsection{Proof of Prop.~\ref{prop:finite-completeness}}
\label{app:finite-completeness}
For any probability distribution $p\in\Delta(\Ftwo^n)$, the zeroth Walsh coefficient is
\begin{align}
\wh p(\boldsymbol{0})=\sum_{\bs}p(\bs)=1.
\end{align}
If $Q_{t_\star}=\Ftwo^n\setminus\{\boldsymbol{0}\}$ and $p\in\cF_{t_\star}$, then every Walsh
coefficient of $p$ agrees with the corresponding Walsh coefficient of $p_\rho$: the nonzero
coefficients agree by definition of $\cF_{t_\star}$, while the zeroth coefficient agrees by
normalization. Walsh--Hadamard inversion gives $p=p_\rho$, so
\begin{align}
\cF_{t_\star}=\{p_\rho\}.
\end{align}
For every $t\ge t_\star$, monotonicity gives $\cF_t\subseteq\cF_{t_\star}$, and the true syndrome
distribution always satisfies the queried exact constraints, so $p_\rho\in\cF_t$. Hence
$\cF_t=\{p_\rho\}$ and
\begin{align}
L_t=U_t=p_\rho(\boldsymbol{0})=F(\rho,\psi).
\end{align}
Finally, if $\bigcup_{t\ge 1}Q_t=\Ftwo^n\setminus\{\boldsymbol{0}\}$, then the right-hand side is
finite, so there is a finite round $t_\star$ by which every nonzero label has appeared. The preceding
argument gives finite-time convergence.

\appsubsection{Proof of Prop.~\ref{prop:query-complexity-lower-bound}}
\label{app:lowerbound}
Assume an adaptive generator-only protocol has queried a set
$Q\subseteq\Ftwo^n\setminus\{\boldsymbol{0}\}$ of nontrivial stabilizer labels, and suppose that
some nonzero direction $\bv\notin Q$ remains unqueried. For any $\eta\in(0,1]$, define
\begin{align}
p_\pm(\bs):=2^{-n}\bigl(1\pm \eta(-1)^{\bv\cdot \bs}\bigr),
\qquad \bs\in\Ftwo^n.
\end{align}
These are valid probability distributions because
$1\pm \eta(-1)^{\bv\cdot \bs}\in[0,2]$ and the Walsh character
$(-1)^{\bv\cdot\bs}$ has zero average over $\bs$. By Walsh orthogonality,
\begin{align}
\wh p_\pm(\bu)
&=\sum_{\bs\in\Ftwo^n}(-1)^{\bu\cdot\bs}2^{-n}
\bigl(1\pm\eta(-1)^{\bv\cdot\bs}\bigr)\nonumber\\
&=\mathbbm{1}[\bu=\boldsymbol{0}]\pm\eta\,\mathbbm{1}[\bu=\bv].
\end{align}
Therefore $\wh p_+(\bu)=\wh p_-(\bu)$ for every $\bu\in Q$, because $\bv\notin Q$ and all queried
labels are nonzero. The protocol receives the same data from $p_+$ and $p_-$, but
\begin{align}
p_+(\boldsymbol{0})=2^{-n}(1+\eta),
\qquad
p_-(\boldsymbol{0})=2^{-n}(1-\eta),
\end{align}
so the target fidelities differ. No protocol can always output the exact fidelity before every
nonzero character has been queried. Since one gauge contributes at most $n$ nontrivial stabilizer
labels, worst-case exact determination requires at least
\begin{align}
\left\lceil\frac{2^n-1}{n}\right\rceil
\end{align}
full gauge rounds.

\appsubsection{Proof of Prop.~\ref{prop:coverage-rate}}
\label{app:coverage-rate}
Let $p,q\in\cF_t$. Walsh--Hadamard inversion at the zero syndrome gives
\begin{align}
p(\boldsymbol{0})-q(\boldsymbol{0})
=2^{-n}\sum_{\bu\in\Ftwo^n}\bigl(\wh p(\bu)-\wh q(\bu)\bigr).
\end{align}
The term $\bu=\boldsymbol{0}$ vanishes because both $p$ and $q$ are probability distributions, and every term with $\bu\in Q_t$ vanishes because $p$ and $q$ satisfy the same queried Walsh constraints. Hence
\begin{align}
p(\boldsymbol{0})-q(\boldsymbol{0})
=2^{-n}\sum_{\bu\in (\Ftwo^n\setminus\{\boldsymbol{0}\})\setminus Q_t}\bigl(\wh p(\bu)-\wh q(\bu)\bigr).
\end{align}
For every probability distribution $r$, one has $|\wh r(\bu)|\le 1$. Therefore
\begin{align}
|p(\boldsymbol{0})-q(\boldsymbol{0})|
\le 2^{-n}\sum_{\bu\in (\Ftwo^n\setminus\{\boldsymbol{0}\})\setminus Q_t}2
=\frac{2m_t}{2^n}.
\end{align}
Taking $p$ and $q$ to be endpoint witnesses for the maximum and minimum of $p(\boldsymbol{0})$ over $\cF_t$ gives Eq.~\eqref{eq:coverage-width-bound}; the additional upper bound by $1$ is trivial because $W_t$ is a probability interval width.

For the randomized statement, let $I_t(\bu)=\mathbbm{1}[\bu\notin Q_t]$. The conditional coverage assumption implies
\begin{align}
\bE\!\left[I_{t+1}(\bu)\mid \cH_t\right]\le (1-\alpha)I_t(\bu).
\end{align}
Iterating from $Q_0=\emptyset$ gives $\bE I_t(\bu)\le (1-\alpha)^t$ for each $\bu\in\Ftwo^n\setminus\{\boldsymbol{0}\}$. Consequently,
\begin{align}
\bE m_t
=\sum_{\bu\in \Ftwo^n\setminus\{\boldsymbol{0}\}}\bE I_t(\bu)
\le (2^n-1)(1-\alpha)^t.
\end{align}
Combining this with Eq.~\eqref{eq:coverage-width-bound} proves Eq.~\eqref{eq:expected-width-rate}.

If a gauge is sampled uniformly from $\GL(n,2)$, its column set has size $n$ and, by symmetry, contains any fixed nonzero label with probability $n/(2^n-1)$. Thus $\alpha=n/(2^n-1)$. Substitution gives the first inequality in Eq.~\eqref{eq:uniform-random-width-rate}, and the second follows from $1-x\le e^{-x}$.

\appsubsection{Proof of Prop.~\ref{prop:disagreement-mass}}
\label{app:disagreement-mass}
Applying Walsh--Hadamard inversion to the endpoint witnesses gives
\begin{align}
W_t
= p_t^+(\boldsymbol{0})-p_t^-(\boldsymbol{0})
= 2^{-n}\sum_{\bu\in\Ftwo^n}
\bigl(\wh p_t^+(\bu)-\wh p_t^-(\bu)\bigr).
\end{align}
The zeroth term vanishes because both endpoint witnesses are probability distributions, and every
queried term vanishes because both witnesses lie in $\cF_t$. Hence
\begin{align}
W_t
=2^{-n}
\sum_{\bu\in(\Ftwo^n\setminus\{\boldsymbol{0}\})\setminus Q_t}
\bigl(\wh p_t^+(\bu)-\wh p_t^-(\bu)\bigr)
\le 2^{-n}D_t .
\end{align}
Since $D_t\le m_t\Delta_t$, Eq.~\eqref{eq:disagreement-mass-bound} follows. If $W_t>0$, this gives
$\Delta_t\ge 2^nW_t/m_t>0$.

Let $\bu_\star$ be an unqueried label attaining $\Delta_t$. Any nonzero vector in $\Ftwo^n$ can be
extended to a vector space basis, so there exists a gauge whose column set contains $\bu_\star$.
That gauge belongs to $\Gamma_t$, because it contains the unqueried label $\bu_\star$. All
disagreement scores are nonnegative, so the maximum weight gauge selected by
Eq.~\eqref{eq:next-basis} has score at least $d_t(\bu_\star)=\Delta_t$. This proves
Eq.~\eqref{eq:gauge-score-lower-bound}.

\appsubsection{Proof of Prop.~\ref{prop:structured-noise}}
\label{app:structured}
Let $p=p_{\bs_0,V}$ where $V\le \Ftwo^n$ has dimension $r$ and codimension $d=n-r$. For any $\bu\in\Ftwo^n$,
\begin{align}
\wh p(\bu)
=\sum_{\bs\in\Ftwo^n}(-1)^{\bu\cdot \bs}p(\bs)
=2^{-r}\sum_{\bv\in V}(-1)^{\bu\cdot(\bs_0+\bv)}
=(-1)^{\bu\cdot \bs_0}\,2^{-r}\sum_{\bv\in V}(-1)^{\bu\cdot \bv}.
\end{align}
If $\bu\in V^\perp$, every term in the last sum equals $1$, so $\wh p(\bu)=(-1)^{\bu\cdot \bs_0}$. If $\bu\notin V^\perp$, then the character $\bv\mapsto (-1)^{\bu\cdot \bv}$ is nontrivial on $V$ and sums to $0$. Therefore
\begin{align}
\label{eq:affine-fourier}
\wh p(\bu)=(-1)^{\bu\cdot \bs_0}\mathbbm{1}[\bu\in V^\perp].
\end{align}
Suppose first that $Q_t$ contains a basis $\bu_1,\cdots,\bu_d$ of $V^\perp$. For any feasible
syndrome distribution $q\in\cF_t$, the constraints
\begin{align}
\wh q(\bu_j)=\wh p(\bu_j)=(-1)^{\bu_j\cdot\bs_0},\qquad j=1,\cdots,d,
\end{align}
force the character $(-1)^{\bu_j\cdot\bs}$ to be constant on the support of $q$. Hence every
feasible $q$ is supported on
\begin{align}
\{\bs\in\Ftwo^n:\bu_j\cdot\bs=\bu_j\cdot\bs_0\ \forall j\}
=\bs_0+V.
\end{align}
If $\bs_0\notin V$, then $\boldsymbol{0}\notin\bs_0+V$, so every feasible $q$ has
$q(\boldsymbol{0})=0$. Therefore $L_t=U_t=0=F(\rho,\psi)$, and
Alg.~\ref{alg:adaptive} terminates by round $t$.

It remains to consider the case $\bs_0\in V$. Then the preceding support constraint reduces every
feasible $q$ to a distribution on $V$. Assume in addition that $Q_t$ contains one representative of
every nonzero coset of $\Ftwo^n/V^\perp$. The restrictions of these representatives to $V$ are
exactly all nonzero Walsh characters on $V$, and Eq.~\eqref{eq:affine-fourier} gives their true
expectation value as $0$. Thus every feasible $q$ supported on $V$ has all nonzero Walsh
coefficients on $V$ equal to zero. Walsh--Hadamard inversion on the subspace $V$ then gives
$q(\bs)=2^{-r}$ for every $\bs\in V$. In particular,
$q(\boldsymbol{0})=2^{-r}=F(\rho,\psi)$ for every feasible $q$, so
$L_t=U_t=F(\rho,\psi)$ and Alg.~\ref{alg:adaptive} terminates by round $t$.

The label count follows because a basis of $V^\perp$ has $d$ labels and
$\Ftwo^n/V^\perp$ has $2^r-1$ nonzero cosets.


\begin{thebibliography}{33}%
\makeatletter
\providecommand \@ifxundefined [1]{%
 \@ifx{#1\undefined}
}%
\providecommand \@ifnum [1]{%
 \ifnum #1\expandafter \@firstoftwo
 \else \expandafter \@secondoftwo
 \fi
}%
\providecommand \@ifx [1]{%
 \ifx #1\expandafter \@firstoftwo
 \else \expandafter \@secondoftwo
 \fi
}%
\providecommand \natexlab [1]{#1}%
\providecommand \enquote  [1]{``#1''}%
\providecommand \bibnamefont  [1]{#1}%
\providecommand \bibfnamefont [1]{#1}%
\providecommand \citenamefont [1]{#1}%
\providecommand \href@noop [0]{\@secondoftwo}%
\providecommand \href [0]{\begingroup \@sanitize@url \@href}%
\providecommand \@href[1]{\@@startlink{#1}\@@href}%
\providecommand \@@href[1]{\endgroup#1\@@endlink}%
\providecommand \@sanitize@url [0]{\catcode `\\12\catcode `\$12\catcode `\&12\catcode `\#12\catcode `\^12\catcode `\_12\catcode `\%12\relax}%
\providecommand \@@startlink[1]{}%
\providecommand \@@endlink[0]{}%
\providecommand \url  [0]{\begingroup\@sanitize@url \@url }%
\providecommand \@url [1]{\endgroup\@href {#1}{\urlprefix }}%
\providecommand \urlprefix  [0]{URL }%
\providecommand \Eprint [0]{\href }%
\providecommand \doibase [0]{https://doi.org/}%
\providecommand \selectlanguage [0]{\@gobble}%
\providecommand \bibinfo  [0]{\@secondoftwo}%
\providecommand \bibfield  [0]{\@secondoftwo}%
\providecommand \translation [1]{[#1]}%
\providecommand \BibitemOpen [0]{}%
\providecommand \bibitemStop [0]{}%
\providecommand \bibitemNoStop [0]{.\EOS\space}%
\providecommand \EOS [0]{\spacefactor3000\relax}%
\providecommand \BibitemShut  [1]{\csname bibitem#1\endcsname}%
\let\auto@bib@innerbib\@empty
\bibitem [{\citenamefont {Gottesman}(1997)}]{Gottesman1997}%
  \BibitemOpen
  \bibfield  {author} {\bibinfo {author} {\bibfnamefont {D.}~\bibnamefont {Gottesman}},\ }\emph {\bibinfo {title} {Stabilizer Codes and Quantum Error Correction}},\ \href@noop {} {Ph.D. thesis},\ \bibinfo  {school} {California Institute of Technology} (\bibinfo {year} {1997}),\ \Eprint {https://arxiv.org/abs/quant-ph/9705052} {arXiv:quant-ph/9705052} \BibitemShut {NoStop}%
\bibitem [{\citenamefont {Terhal}(2015)}]{Terhal2015}%
  \BibitemOpen
  \bibfield  {author} {\bibinfo {author} {\bibfnamefont {B.~M.}\ \bibnamefont {Terhal}},\ }\href {https://doi.org/10.1103/RevModPhys.87.307} {\bibfield  {journal} {\bibinfo  {journal} {Rev. Mod. Phys.}\ }\textbf {\bibinfo {volume} {87}},\ \bibinfo {pages} {307–346} (\bibinfo {year} {2015})}\BibitemShut {NoStop}%
\bibitem [{\citenamefont {Raussendorf}\ and\ \citenamefont {Briegel}(2001)}]{RaussendorfBriegel2001}%
  \BibitemOpen
  \bibfield  {author} {\bibinfo {author} {\bibfnamefont {R.}~\bibnamefont {Raussendorf}}\ and\ \bibinfo {author} {\bibfnamefont {H.~J.}\ \bibnamefont {Briegel}},\ }\href {https://doi.org/10.1038/35051009} {\bibfield  {journal} {\bibinfo  {journal} {Nature}\ }\textbf {\bibinfo {volume} {409}},\ \bibinfo {pages} {46–52} (\bibinfo {year} {2001})}\BibitemShut {NoStop}%
\bibitem [{\citenamefont {Walther}\ \emph {et~al.}(2005)\citenamefont {Walther}, \citenamefont {Resch}, \citenamefont {Rudolph}, \citenamefont {Schenck}, \citenamefont {Weinfurter}, \citenamefont {Vedral}, \citenamefont {Aspelmeyer},\ and\ \citenamefont {Zeilinger}}]{Walther2005}%
  \BibitemOpen
  \bibfield  {author} {\bibinfo {author} {\bibfnamefont {P.}~\bibnamefont {Walther}}, \bibinfo {author} {\bibfnamefont {K.~J.}\ \bibnamefont {Resch}}, \bibinfo {author} {\bibfnamefont {T.}~\bibnamefont {Rudolph}}, \bibinfo {author} {\bibfnamefont {E.}~\bibnamefont {Schenck}}, \bibinfo {author} {\bibfnamefont {H.}~\bibnamefont {Weinfurter}}, \bibinfo {author} {\bibfnamefont {V.}~\bibnamefont {Vedral}}, \bibinfo {author} {\bibfnamefont {M.}~\bibnamefont {Aspelmeyer}},\ and\ \bibinfo {author} {\bibfnamefont {A.}~\bibnamefont {Zeilinger}},\ }\href {https://doi.org/10.1038/nature03347} {\bibfield  {journal} {\bibinfo  {journal} {Nature}\ }\textbf {\bibinfo {volume} {434}},\ \bibinfo {pages} {169–176} (\bibinfo {year} {2005})}\BibitemShut {NoStop}%
\bibitem [{\citenamefont {Barends}\ \emph {et~al.}(2014)\citenamefont {Barends} \emph {et~al.}}]{Barends2014}%
  \BibitemOpen
  \bibfield  {author} {\bibinfo {author} {\bibfnamefont {R.}~\bibnamefont {Barends}} \emph {et~al.},\ }\href {https://doi.org/10.1038/nature13171} {\bibfield  {journal} {\bibinfo  {journal} {Nature}\ }\textbf {\bibinfo {volume} {508}},\ \bibinfo {pages} {500–503} (\bibinfo {year} {2014})}\BibitemShut {NoStop}%
\bibitem [{\citenamefont {Kelly}\ \emph {et~al.}(2015)\citenamefont {Kelly} \emph {et~al.}}]{Kelly2015}%
  \BibitemOpen
  \bibfield  {author} {\bibinfo {author} {\bibfnamefont {J.}~\bibnamefont {Kelly}} \emph {et~al.},\ }\href {https://doi.org/10.1038/nature14270} {\bibfield  {journal} {\bibinfo  {journal} {Nature}\ }\textbf {\bibinfo {volume} {519}},\ \bibinfo {pages} {66–69} (\bibinfo {year} {2015})}\BibitemShut {NoStop}%
\bibitem [{\citenamefont {Cao}\ \emph {et~al.}(2023)\citenamefont {Cao} \emph {et~al.}}]{Cao2023}%
  \BibitemOpen
  \bibfield  {author} {\bibinfo {author} {\bibfnamefont {S.}~\bibnamefont {Cao}} \emph {et~al.},\ }\href {https://doi.org/10.1038/s41586-023-06195-1} {\bibfield  {journal} {\bibinfo  {journal} {Nature}\ }\textbf {\bibinfo {volume} {619}},\ \bibinfo {pages} {738–742} (\bibinfo {year} {2023})}\BibitemShut {NoStop}%
\bibitem [{\citenamefont {Jiang}\ \emph {et~al.}(2026)\citenamefont {Jiang} \emph {et~al.}}]{Jiang2026}%
  \BibitemOpen
  \bibfield  {author} {\bibinfo {author} {\bibfnamefont {T.}~\bibnamefont {Jiang}} \emph {et~al.},\ }\href {https://doi.org/10.1038/s41567-026-03179-6} {\bibfield  {journal} {\bibinfo  {journal} {Nature Phys.}\ }\textbf {\bibinfo {volume} {22}},\ \bibinfo {pages} {430–438} (\bibinfo {year} {2026})}\BibitemShut {NoStop}%
\bibitem [{\citenamefont {Satzinger}\ \emph {et~al.}(2021)\citenamefont {Satzinger} \emph {et~al.}}]{Satzinger2021}%
  \BibitemOpen
  \bibfield  {author} {\bibinfo {author} {\bibfnamefont {K.~J.}\ \bibnamefont {Satzinger}} \emph {et~al.},\ }\href {https://doi.org/10.1126/science.abi8378} {\bibfield  {journal} {\bibinfo  {journal} {Science}\ }\textbf {\bibinfo {volume} {374}},\ \bibinfo {pages} {1237–1241} (\bibinfo {year} {2021})}\BibitemShut {NoStop}%
\bibitem [{\citenamefont {{Google Quantum AI}}(2023)}]{GoogleQuantumAI2023}%
  \BibitemOpen
  \bibfield  {author} {\bibinfo {author} {\bibnamefont {{Google Quantum AI}}},\ }\href {https://doi.org/10.1038/s41586-022-05434-1} {\bibfield  {journal} {\bibinfo  {journal} {Nature}\ }\textbf {\bibinfo {volume} {614}},\ \bibinfo {pages} {676–681} (\bibinfo {year} {2023})}\BibitemShut {NoStop}%
\bibitem [{\citenamefont {Egan}\ \emph {et~al.}(2021)\citenamefont {Egan} \emph {et~al.}}]{Egan2021}%
  \BibitemOpen
  \bibfield  {author} {\bibinfo {author} {\bibfnamefont {L.}~\bibnamefont {Egan}} \emph {et~al.},\ }\href {https://doi.org/10.1038/s41586-021-03928-y} {\bibfield  {journal} {\bibinfo  {journal} {Nature}\ }\textbf {\bibinfo {volume} {598}},\ \bibinfo {pages} {281–286} (\bibinfo {year} {2021})}\BibitemShut {NoStop}%
\bibitem [{\citenamefont {Bluvstein}\ \emph {et~al.}(2024)\citenamefont {Bluvstein} \emph {et~al.}}]{Bluvstein2024}%
  \BibitemOpen
  \bibfield  {author} {\bibinfo {author} {\bibfnamefont {D.}~\bibnamefont {Bluvstein}} \emph {et~al.},\ }\href {https://doi.org/10.1038/s41586-023-06927-3} {\bibfield  {journal} {\bibinfo  {journal} {Nature}\ }\textbf {\bibinfo {volume} {626}},\ \bibinfo {pages} {58–65} (\bibinfo {year} {2024})}\BibitemShut {NoStop}%
\bibitem [{\citenamefont {Bluvstein}\ \emph {et~al.}(2025)\citenamefont {Bluvstein} \emph {et~al.}}]{Bluvstein2025}%
  \BibitemOpen
  \bibfield  {author} {\bibinfo {author} {\bibfnamefont {D.}~\bibnamefont {Bluvstein}} \emph {et~al.},\ }\href {https://doi.org/10.1038/s41586-025-09848-5} {\bibfield  {journal} {\bibinfo  {journal} {Nature}\ }\textbf {\bibinfo {volume} {649}},\ \bibinfo {pages} {39–46} (\bibinfo {year} {2025})}\BibitemShut {NoStop}%
\bibitem [{\citenamefont {James}\ \emph {et~al.}(2001)\citenamefont {James}, \citenamefont {Kwiat}, \citenamefont {Munro},\ and\ \citenamefont {White}}]{James2001}%
  \BibitemOpen
  \bibfield  {author} {\bibinfo {author} {\bibfnamefont {D.~F.~V.}\ \bibnamefont {James}}, \bibinfo {author} {\bibfnamefont {P.~G.}\ \bibnamefont {Kwiat}}, \bibinfo {author} {\bibfnamefont {W.~J.}\ \bibnamefont {Munro}},\ and\ \bibinfo {author} {\bibfnamefont {A.~G.}\ \bibnamefont {White}},\ }\href {https://doi.org/10.1103/PhysRevA.64.052312} {\bibfield  {journal} {\bibinfo  {journal} {Phys. Rev. A}\ }\textbf {\bibinfo {volume} {64}},\ \bibinfo {pages} {052312} (\bibinfo {year} {2001})}\BibitemShut {NoStop}%
\bibitem [{\citenamefont {Altepeter}\ \emph {et~al.}(2005)\citenamefont {Altepeter}, \citenamefont {Jeffrey},\ and\ \citenamefont {Kwiat}}]{Altepeter2005}%
  \BibitemOpen
  \bibfield  {author} {\bibinfo {author} {\bibfnamefont {J.~B.}\ \bibnamefont {Altepeter}}, \bibinfo {author} {\bibfnamefont {E.~R.}\ \bibnamefont {Jeffrey}},\ and\ \bibinfo {author} {\bibfnamefont {P.~G.}\ \bibnamefont {Kwiat}},\ }in\ \href {https://doi.org/10.1016/S1049-250X(05)52003-2} {\emph {\bibinfo {booktitle} {Advances in Atomic, Molecular, and Optical Physics}}},\ Vol.~\bibinfo {volume} {52}\ (\bibinfo  {publisher} {Academic Press},\ \bibinfo {year} {2005})\ p.\ \bibinfo {pages} {105–159}\BibitemShut {NoStop}%
\bibitem [{\citenamefont {Gross}\ \emph {et~al.}(2010)\citenamefont {Gross}, \citenamefont {Liu}, \citenamefont {Flammia}, \citenamefont {Becker},\ and\ \citenamefont {Eisert}}]{Gross2010}%
  \BibitemOpen
  \bibfield  {author} {\bibinfo {author} {\bibfnamefont {D.}~\bibnamefont {Gross}}, \bibinfo {author} {\bibfnamefont {Y.-K.}\ \bibnamefont {Liu}}, \bibinfo {author} {\bibfnamefont {S.~T.}\ \bibnamefont {Flammia}}, \bibinfo {author} {\bibfnamefont {S.}~\bibnamefont {Becker}},\ and\ \bibinfo {author} {\bibfnamefont {J.}~\bibnamefont {Eisert}},\ }\href {https://doi.org/10.1103/PhysRevLett.105.150401} {\bibfield  {journal} {\bibinfo  {journal} {Phys. Rev. Lett.}\ }\textbf {\bibinfo {volume} {105}},\ \bibinfo {pages} {150401} (\bibinfo {year} {2010})}\BibitemShut {NoStop}%
\bibitem [{\citenamefont {Cramer}\ \emph {et~al.}(2010)\citenamefont {Cramer} \emph {et~al.}}]{Cramer2010}%
  \BibitemOpen
  \bibfield  {author} {\bibinfo {author} {\bibfnamefont {M.}~\bibnamefont {Cramer}} \emph {et~al.},\ }\href {https://doi.org/10.1038/ncomms1147} {\bibfield  {journal} {\bibinfo  {journal} {Nature Commun.}\ }\textbf {\bibinfo {volume} {1}},\ \bibinfo {pages} {149} (\bibinfo {year} {2010})}\BibitemShut {NoStop}%
\bibitem [{\citenamefont {Hayashi}\ and\ \citenamefont {Morimae}(2015)}]{HayashiMorimae2015}%
  \BibitemOpen
  \bibfield  {author} {\bibinfo {author} {\bibfnamefont {M.}~\bibnamefont {Hayashi}}\ and\ \bibinfo {author} {\bibfnamefont {T.}~\bibnamefont {Morimae}},\ }\href {https://doi.org/10.1103/PhysRevLett.115.220502} {\bibfield  {journal} {\bibinfo  {journal} {Phys. Rev. Lett.}\ }\textbf {\bibinfo {volume} {115}},\ \bibinfo {pages} {220502} (\bibinfo {year} {2015})}\BibitemShut {NoStop}%
\bibitem [{\citenamefont {Dangniam}\ \emph {et~al.}(2020)\citenamefont {Dangniam}, \citenamefont {Han},\ and\ \citenamefont {Zhu}}]{DangniamHanZhu2020}%
  \BibitemOpen
  \bibfield  {author} {\bibinfo {author} {\bibfnamefont {N.}~\bibnamefont {Dangniam}}, \bibinfo {author} {\bibfnamefont {Y.-G.}\ \bibnamefont {Han}},\ and\ \bibinfo {author} {\bibfnamefont {H.}~\bibnamefont {Zhu}},\ }\href {https://doi.org/10.1103/PhysRevResearch.2.043323} {\bibfield  {journal} {\bibinfo  {journal} {Phys. Rev. Research}\ }\textbf {\bibinfo {volume} {2}},\ \bibinfo {pages} {043323} (\bibinfo {year} {2020})}\BibitemShut {NoStop}%
\bibitem [{\citenamefont {Li}\ \emph {et~al.}(2020)\citenamefont {Li}, \citenamefont {Han},\ and\ \citenamefont {Zhu}}]{LiHanZhu2020}%
  \BibitemOpen
  \bibfield  {author} {\bibinfo {author} {\bibfnamefont {Z.}~\bibnamefont {Li}}, \bibinfo {author} {\bibfnamefont {Y.-G.}\ \bibnamefont {Han}},\ and\ \bibinfo {author} {\bibfnamefont {H.}~\bibnamefont {Zhu}},\ }\href {https://doi.org/10.1103/PhysRevApplied.13.054002} {\bibfield  {journal} {\bibinfo  {journal} {Phys. Rev. Applied}\ }\textbf {\bibinfo {volume} {13}},\ \bibinfo {pages} {054002} (\bibinfo {year} {2020})}\BibitemShut {NoStop}%
\bibitem [{\citenamefont {Chen}\ \emph {et~al.}(2025)\citenamefont {Chen}, \citenamefont {Xie}, \citenamefont {Xu},\ and\ \citenamefont {Wang}}]{ChenXieXuWang2025}%
  \BibitemOpen
  \bibfield  {author} {\bibinfo {author} {\bibfnamefont {S.}~\bibnamefont {Chen}}, \bibinfo {author} {\bibfnamefont {W.}~\bibnamefont {Xie}}, \bibinfo {author} {\bibfnamefont {P.}~\bibnamefont {Xu}},\ and\ \bibinfo {author} {\bibfnamefont {K.}~\bibnamefont {Wang}},\ }\href {https://doi.org/10.1103/PhysRevResearch.7.013003} {\bibfield  {journal} {\bibinfo  {journal} {Phys. Rev. Research}\ }\textbf {\bibinfo {volume} {7}},\ \bibinfo {pages} {013003} (\bibinfo {year} {2025})}\BibitemShut {NoStop}%
\bibitem [{\citenamefont {Zheng}\ \emph {et~al.}(2026)\citenamefont {Zheng}, \citenamefont {Yu}, \citenamefont {Zhang}, \citenamefont {Xu},\ and\ \citenamefont {Wang}}]{ZhengYuZhangXuWang2026}%
  \BibitemOpen
  \bibfield  {author} {\bibinfo {author} {\bibfnamefont {C.}~\bibnamefont {Zheng}}, \bibinfo {author} {\bibfnamefont {X.}~\bibnamefont {Yu}}, \bibinfo {author} {\bibfnamefont {Z.}~\bibnamefont {Zhang}}, \bibinfo {author} {\bibfnamefont {P.}~\bibnamefont {Xu}},\ and\ \bibinfo {author} {\bibfnamefont {K.}~\bibnamefont {Wang}},\ }\href {https://doi.org/10.1088/2058-9565/ae723b} {\bibfield  {journal} {\bibinfo  {journal} {Quantum Sci. Technol.}\ }\textbf {\bibinfo {volume} {X}},\ \bibinfo {pages} {XXXXXX} (\bibinfo {year} {2026})},\ \bibinfo {note} {accepted Manuscript online; arXiv:2409.19699}\BibitemShut {NoStop}%
\bibitem [{\citenamefont {Flammia}\ and\ \citenamefont {Liu}(2011)}]{FlammiaLiu2011}%
  \BibitemOpen
  \bibfield  {author} {\bibinfo {author} {\bibfnamefont {S.~T.}\ \bibnamefont {Flammia}}\ and\ \bibinfo {author} {\bibfnamefont {Y.-K.}\ \bibnamefont {Liu}},\ }\href {https://doi.org/10.1103/PhysRevLett.106.230501} {\bibfield  {journal} {\bibinfo  {journal} {Phys. Rev. Lett.}\ }\textbf {\bibinfo {volume} {106}},\ \bibinfo {pages} {230501} (\bibinfo {year} {2011})}\BibitemShut {NoStop}%
\bibitem [{\citenamefont {da~Silva}\ \emph {et~al.}(2011)\citenamefont {da~Silva}, \citenamefont {Landon-Cardinal},\ and\ \citenamefont {Poulin}}]{daSilvaLandonCardinalPoulin2011}%
  \BibitemOpen
  \bibfield  {author} {\bibinfo {author} {\bibfnamefont {M.~P.}\ \bibnamefont {da~Silva}}, \bibinfo {author} {\bibfnamefont {O.}~\bibnamefont {Landon-Cardinal}},\ and\ \bibinfo {author} {\bibfnamefont {D.}~\bibnamefont {Poulin}},\ }\href {https://doi.org/10.1103/PhysRevLett.107.210404} {\bibfield  {journal} {\bibinfo  {journal} {Phys. Rev. Lett.}\ }\textbf {\bibinfo {volume} {107}},\ \bibinfo {pages} {210404} (\bibinfo {year} {2011})}\BibitemShut {NoStop}%
\bibitem [{\citenamefont {Kalev}\ \emph {et~al.}(2019)\citenamefont {Kalev}, \citenamefont {Kyrillidis},\ and\ \citenamefont {Linke}}]{Kalev2019}%
  \BibitemOpen
  \bibfield  {author} {\bibinfo {author} {\bibfnamefont {A.}~\bibnamefont {Kalev}}, \bibinfo {author} {\bibfnamefont {A.}~\bibnamefont {Kyrillidis}},\ and\ \bibinfo {author} {\bibfnamefont {N.~M.}\ \bibnamefont {Linke}},\ }\href {https://doi.org/10.1103/PhysRevA.99.042337} {\bibfield  {journal} {\bibinfo  {journal} {Phys. Rev. A}\ }\textbf {\bibinfo {volume} {99}},\ \bibinfo {pages} {042337} (\bibinfo {year} {2019})}\BibitemShut {NoStop}%
\bibitem [{\citenamefont {Kelley}(1960)}]{Kelley1960}%
  \BibitemOpen
  \bibfield  {author} {\bibinfo {author} {\bibfnamefont {J.~E.}\ \bibnamefont {Kelley}},\ }\href {https://doi.org/10.1137/0108053} {\bibfield  {journal} {\bibinfo  {journal} {J. Soc. Indust. Appl. Math.}\ }\textbf {\bibinfo {volume} {8}},\ \bibinfo {pages} {703–712} (\bibinfo {year} {1960})}\BibitemShut {NoStop}%
\bibitem [{\citenamefont {Bertsimas}\ and\ \citenamefont {Tsitsiklis}(1997)}]{BertsimasTsitsiklis1997}%
  \BibitemOpen
  \bibfield  {author} {\bibinfo {author} {\bibfnamefont {D.}~\bibnamefont {Bertsimas}}\ and\ \bibinfo {author} {\bibfnamefont {J.~N.}\ \bibnamefont {Tsitsiklis}},\ }\href@noop {} {\emph {\bibinfo {title} {Introduction to Linear Optimization}}}\ (\bibinfo  {publisher} {Athena Scientific},\ \bibinfo {address} {Belmont, MA},\ \bibinfo {year} {1997})\BibitemShut {NoStop}%
\bibitem [{\citenamefont {Desaulniers}\ \emph {et~al.}(2005)\citenamefont {Desaulniers}, \citenamefont {Desrosiers},\ and\ \citenamefont {Solomon}}]{Desaulniers2005}%
  \BibitemOpen
  \bibinfo {editor} {\bibfnamefont {G.}~\bibnamefont {Desaulniers}}, \bibinfo {editor} {\bibfnamefont {J.}~\bibnamefont {Desrosiers}},\ and\ \bibinfo {editor} {\bibfnamefont {M.~M.}\ \bibnamefont {Solomon}},\ eds.,\ \href@noop {} {\emph {\bibinfo {title} {Column Generation}}}\ (\bibinfo  {publisher} {Springer},\ \bibinfo {address} {New York},\ \bibinfo {year} {2005})\BibitemShut {NoStop}%
\bibitem [{\citenamefont {L{\"u}bbecke}\ and\ \citenamefont {Desrosiers}(2005)}]{LuebbeckeDesrosiers2005}%
  \BibitemOpen
  \bibfield  {author} {\bibinfo {author} {\bibfnamefont {M.~E.}\ \bibnamefont {L{\"u}bbecke}}\ and\ \bibinfo {author} {\bibfnamefont {J.}~\bibnamefont {Desrosiers}},\ }\href {https://doi.org/10.1287/opre.1050.0234} {\bibfield  {journal} {\bibinfo  {journal} {Oper. Res.}\ }\textbf {\bibinfo {volume} {53}},\ \bibinfo {pages} {1007–1023} (\bibinfo {year} {2005})}\BibitemShut {NoStop}%
\bibitem [{\citenamefont {Rocchetto}(2018)}]{Rocchetto2018}%
  \BibitemOpen
  \bibfield  {author} {\bibinfo {author} {\bibfnamefont {A.}~\bibnamefont {Rocchetto}},\ }\href {https://doi.org/10.5555/3524938.3524974} {\bibfield  {journal} {\bibinfo  {journal} {Quantum Inf. Comput.}\ }\textbf {\bibinfo {volume} {18}},\ \bibinfo {pages} {541–552} (\bibinfo {year} {2018})}\BibitemShut {NoStop}%
\bibitem [{\citenamefont {Montanaro}(2017)}]{Montanaro2017}%
  \BibitemOpen
  \bibfield  {author} {\bibinfo {author} {\bibfnamefont {A.}~\bibnamefont {Montanaro}},\ }\href@noop {} {\bibinfo {title} {Learning stabilizer states by bell sampling}} (\bibinfo {year} {2017}),\ \Eprint {https://arxiv.org/abs/1707.04012} {arXiv:1707.04012} \BibitemShut {NoStop}%
\bibitem [{\citenamefont {Aaronson}\ and\ \citenamefont {Gottesman}(2004)}]{AaronsonGottesman2004}%
  \BibitemOpen
  \bibfield  {author} {\bibinfo {author} {\bibfnamefont {S.}~\bibnamefont {Aaronson}}\ and\ \bibinfo {author} {\bibfnamefont {D.}~\bibnamefont {Gottesman}},\ }\href {https://doi.org/10.1103/PhysRevA.70.052328} {\bibfield  {journal} {\bibinfo  {journal} {Phys. Rev. A}\ }\textbf {\bibinfo {volume} {70}},\ \bibinfo {pages} {052328} (\bibinfo {year} {2004})}\BibitemShut {NoStop}%
\bibitem [{\citenamefont {Lange}\ \emph {et~al.}(2023)\citenamefont {Lange}, \citenamefont {Kebri{\v{c}}}, \citenamefont {Buser}, \citenamefont {Schollw{\"o}ck}, \citenamefont {Grusdt},\ and\ \citenamefont {Bohrdt}}]{Lange2023}%
  \BibitemOpen
  \bibfield  {author} {\bibinfo {author} {\bibfnamefont {H.}~\bibnamefont {Lange}}, \bibinfo {author} {\bibfnamefont {M.}~\bibnamefont {Kebri{\v{c}}}}, \bibinfo {author} {\bibfnamefont {M.}~\bibnamefont {Buser}}, \bibinfo {author} {\bibfnamefont {U.}~\bibnamefont {Schollw{\"o}ck}}, \bibinfo {author} {\bibfnamefont {F.}~\bibnamefont {Grusdt}},\ and\ \bibinfo {author} {\bibfnamefont {A.}~\bibnamefont {Bohrdt}},\ }\href {https://doi.org/10.22331/q-2023-10-19-1129} {\bibfield  {journal} {\bibinfo  {journal} {Quantum}\ }\textbf {\bibinfo {volume} {7}},\ \bibinfo {pages} {1129} (\bibinfo {year} {2023})}\BibitemShut {NoStop}%
\end{thebibliography}
\end{document}